\documentclass[structabstract]{aa}  
\usepackage{amsmath} 		
\usepackage{graphicx}
\usepackage{float}
\usepackage{subfig}
\usepackage{verbatim}
\usepackage{epstopdf}

\usepackage{txfonts}
\usepackage{natbib}
\bibpunct{(}{)}{;}{a}{}{,} 
\newcommand{\argmin}{\underset{i}{\operatorname{arg\,min}}} 

\begin{document}
   \title{ASPECT: A spectra clustering tool for exploration of large spectral surveys}

   \author{{Aick {in der Au}\inst{1}} 
   \and {Helmut {Meusinger}\inst{1}} 
   \and {Philipp F. {Schalldach}\inst{1}}  
   \and {Mark {Newholm}\inst{1}} }
 
   \institute{{Th\"uringer Landessternwarte Tautenburg, 
   Sternwarte 5, D-07778 Tautenburg, Germany}}
       
   \date{\today}

  \abstract
  {Analysing the empirical output from large surveys is an important challenge 
  in contemporary science. Difficulties arise, in particular, when the database 
  is huge and the properties of the object types to be selected are poorly
  constrained {\em a priori}.}
  {We present the novel, semi-automated clustering tool ASPECT for analysing 
  voluminous archives of spectra.}
  {The heart of the program is a neural 
  network in the form of a Kohonen self-organizing map. The resulting map is 
  designed as an icon map suitable for the inspection by eye. The visual 
  analysis is supported by the option to blend in individual object 
  properties such as 
  redshift, apparent magnitude, or signal-to-noise ratio.
  In addition, the package provides several tools for the
  selection of special spectral types, e.g. local difference maps which 
  reflect the deviations of all spectra from one given input spectrum (real or 
  artificial).  
  }
  {ASPECT is able to produce a two-dimensional topological map of a huge number of spectra.
  The software package enables the user to browse and navigate 
  through a huge data pool and helps them to gain an insight into
  underlying relationships between the spectra and other physical 
  properties and to get the big picture of the entire data set.
  We demonstrate the capability of ASPECT by clustering the entire 
  data pool of $\sim 6\,10^5$ spectra from the Data Release 4 of the Sloan Digital Sky 
  Survey (SDSS). To illustrate the results regarding quality and completeness we 
  track objects from existing catalogues of quasars and carbon stars,
  respectively, and connect the SDSS spectra with morphological information 
  from the GalaxyZoo project. 
   }
  {} 
   
   \maketitle
%


\section{Introduction} 

Astronomy has become a data-intensive science.
Cutting edge research is requiring in particular deep and/or 
wide surveys producing data of unprecedented quality and volume.
The Sloan Digital Sky Survey, \citep[SDSS; ][]{Abazajian209SDSSDR7},
one of the most ambitious and influential astronomical surveys, 
obtained more than $10^6$ spectra of galaxies and quasars.
With the growth of massive data-producing sky surveys such as e.g., 
the Large Synoptic Sky Survey \citep{LSST2009}, 
astronomical research will become even more data-intensive in the 
near future. \citet{Berriman:2011:AAS:2039359.2047483} predict 
a growth rate of 0.5 petabyte of electronically accessible astronomical 
data per year. For example, 
vast and deep surveys using multi-object wide-field
spectrographs, mainly on large aperture telescopes, will be critical 
for attempts to constrain the nature of dark matter, dark energy, and 
the processes of large-scale structure formation 
\citep{Peacock2006,Bell2009,Morales2012}.

Analysing the observational output from a large survey is greatly hindered 
by the sheer size of the data volume. For example, it is desirable to
visualise the output in a big picture that illustrates 
both the diversity of the object types, their differences and similarities, 
but also correlations with certain physical parameters at once.
The selection of the objects of a given spectral type
among hundreds of thousands or even millions of spectra provides another 
problem. In principle, this job can be done by using the output from an 
efficient automated spectroscopic pipeline (e.g., \citet{Stoughton2002SDSSEDR}). 
In the case of particularly interesting,
rare object types with poorly constrained spectral features, however, 
it is not a priori clear if one can trust the pipeline.  
For instance, \citet{Hall2002} had to inspect ~120,000 spectra 
visually to find out 23 broad absorption line quasars with various unusual 
properties.   

We developed a new software tool that is able to organise 
large spectral data pools by means of similarity in a topological map. 
The tool reduces the effort for visual inspection, 
enables easier selection from vast amounts of spectral data, 
and provides a greater picture of the entire data set. 
The approach is based on similarity maps generated using 
self-organising maps (SOM) as developed by \citet{SOM}. 
The SOM technique is an artificial neural network algorithm that uses 
unsupervised learning in order to produce a two-dimensional 
mapping of higher order input data. 

Neural networks have been extensively used in the field of astrophysics, 
primarily for different kinds of classification tasks.  
\citet{Odewahn1992} were the first who applied 
 multilayer perceptrons with backpropagation for an image-based 
discrimination between stars and galaxies. 
\citet{Maehoenen1995} and \citet{Miller1996} 
pioneered the use of SOMs for the same purpose, and
\citet{Andreon2000} continued with work in this field. 
Further, SOMs have been used for classification of light curves 
\citep{brett-2004}, 
gamma-ray bursts \citep{Balastegui2001,Rajaniemi2002GRB}, 
stellar spectra \citep{Jian-qiao01}, 
stellar populations \citep{HernandezPajares1994}, 
and broad absorption line quasar spectra 
\citep{Scaringi2009}
using Learning Vector Quantization, a 
supervised generalisation of SOMs. 
However, 
the application of this 
type of neural network is not only limited to classification tasks. 
For instance, 
\citet{Lesteven96neuralnetworks} 
applied SOMs to organise astronomical publications, 
\citet{Naim1997} 
visualised 
the distribution of galaxies,  
\citet{Way2012} and \citet{Geach2012} 
estimated photometric redshifts, and 
\citet{Torniainen2008} 
analysed 
gigahertz-peaked spectrum (GPS) sources and high frequency peakers (HFP) 
using SOMs in order to find homogeneous groups among the sources. For a 
more complete survey of neural network applications in
astronomy, 
we refer to \citet{Tagliaferri03} 
and \citet{Ciaramella05}. 

In most studies found in the 
literature, 
neural networks have been used 
for some sort of object type classification. 
Therefore, 
a given source 
sample - that consists either of the entire spectra or some associated 
physical properties - is divided into a training and a test data set. 
Then, a small network with a few hundred neurons is trained with the training 
data set and then,
the error rate of the classifier is estimated with 
the second data set. 
Our approach goes beyond this technique since we use the 
network to generate a map that contains every single optical spectrum 
of the source data pool grouped by similarity. 

To achieve this goal, 
our network has to consist of orders of magnitude more 
neurons as compared to networks that are used for classification tasks. 
According to our knowledge, 
common software packages, for 
instance SOM Toolbox for Matlab\footnote{www.cis.hut.fi/somtoolbox}, 
SOM\_PAK\footnote{www.cis.hut.fi/research/som\_pak} or commercial ones 
such as Peltarion\footnote{www.peltarion.com} 
are not capable of handling such large networks, so we 
decided to develop our own software.

This paper presents 
the new software tool ASPECT 
(\underline{A} \underline{SPE}ctra-\underline{C}lustering \underline{T}ool)
for computing and evaluating  
of very large SOMs. The overall process consists of the following steps:
1. Selection and preparation of the spectral data set,
2. preprocessing of the spectra,
3. computing the SOM,
4. visualisation and exploration of the final map.
The last step includes such options as blending selected 
parameters (e.g., coordinates, object type, redshift, redshift error,...) 
over the map, selecting objects from user-defined regions of the
map, identifying objects from an external catalogue,
or searching for spectra of a special type defined by 
a template spectrum.

In the next section, 
we discuss the selection and preparation of our example 
spectral data set. Section 3 describes the used algorithms to generate 
a SOM for $\sim 10^6$ spectra 
and discusses some important implementation details 
and optimisations necessary in order to finish computations in a reasonable 
time frame. 
Then, in Sect. 4, we explain the strength of such a SOM
and show some visualisations of physical properties attached to each 
spectrum. 
Further we demonstrate the application of our approach for searching rare 
spectral types using carbon stars from the catalogues of
\citet{Koester2006} and \citet{Downes2004}.
Finally, in Sect. 5, we shortly discuss two example applications for our SOM:
The search for unusual quasars, and then,
by connecting the SOM with morphological data 
from the Galaxy Zoo project \citep{Lintott2011GalaxyZoo}, 
we illustrate how the achieved results can be combined with external 
data sets from different scientific works.


\section{Database, selection and preparation of the spectral data set}

\subsection{Database: The Sloan Digital Sky Survey}
%
The Sloan Digital Sky Survey \citep[SDSS; ][]{York2000SDSSTechSummary}
is currently one of the most influential surveys in modern astronomy, 
especially in the extragalactic domain. The SDSS provides 
photometric and spectroscopic data for more than one quarter of the sky. 
The survey started in 1998 and has a spectroscopic coverage of 9,274 
square degrees. 
The Data Release 8 \citep{Aihara2011SDSSDR8}
contains spectra of over 1.6\,$10^6$ galaxies, quasars, and stars. 
Imaging and spectroscopic data were taken with the 2.5m telescope at 
Apache Point Observatory, New Mexico. The telescope is equipped with 
two digital fiber-fed spectrographs that can observe 640 spectra at 
once.  Photometric data, processed by automatic imaging pipelines 
\citep{Lupton2001SDSSImaging} was 
later used to select spectra of different object classes (quasars, 
galaxies, luminous red galaxies, stars and serendipitous objects). 
Observed spectra were further automatically processed by a spectroscopic 
pipeline which reduces, corrects, and calibrates the spectra. For each 
spectrum the pipeline determined its spectral type and measured redshift, 
emission, and absorption lines.   

The completion of the original goals of the SDSS and the end of the phase 
known as SDSS-II is marked by the DR7 \citep{Abazajian209SDSSDR7}. 
We started our study on Kohonen mapping of the SDSS spectra at the time
of the DR6 \citep{AdelmanMcCarthy2008SDSSDR6} which contains over 1.2 million 
spectra. The early attempts were 
aimed at a basic understanding of the SOMs rather than analysing the 
complete set of spectra from the latest SDSS data release. We thus used   
the smaller database from the DR4 \citep{AdelmanMcCarthy2006SDSSDR4} with 
about $8\,10^5$ spectra. 
Later on,
we used the $\sim 10^5$ quasar spectra from the 
DR7 for a special application of ASPECT to create a sizeable sample of unusual 
SDSS quasars (\citealp{Meusinger2012}; Sect. 5.1). The aim of the present study, 
namely demonstrating the power and the general properties 
of the SOMs for all types of objects from the SDSS spectroscopic survey,
does not require to involve the complete database from the
last data release. We decided again to use the database from the DR4 simply
in order to reduce the size of the complete map as well as the corresponding 
computing time to a manageable size. The spectra itself were taken from the DR6, 
which operates on an improved spectroscopic pipeline over DR4.
Creating the here presented DR4 map took over 100 days computing time 
on a single workstation\footnote{Intel Core i7 920 at 2.67GHz with 12 GB RAM} 
whereas 
a runtime of nearly 3 years is estimated for the corresponding map from the DR7.
This problem for SDSS DR8 or upcoming data releases could be overcome in two ways. Either by clustering 
multiple smaller maps in parallel, each map on a different workstation, or by distributing the computational workload 
for one large map onto multiple workstations so that computing times are reduced to a manageable length.
Our current software prototype executes already several algorithms in parallel on a single multi-core or multiprocessor machine. 
However distributed computations among multiple computers are not yet supported.

The SDSS spectra cover the wavelength range 
from 3800\AA\ to 9200\AA\ with a resolution of 
$\sim 2000$ and a sampling of $\sim 2.4$ pixels per resolution element. 
Each spectrum is given 
as a \verb|FITS| file and can be identified by the combination of 
its MJD, plate number, and fiber id. In addition to the observed spectrum,
each \verb|FITS| file contains a rich set of parameters and physical 
properties where we are interested in a small fraction only. 

All spectra are stored in the \verb|SpecObjAll| database table.
In order to eliminate useless or undesired spectra, we only took those 
from the \verb|SpecObj| database view. 
According to \citet{Gray2002} duplicate objects, plates for quality assurance, sky 
data or plates that are outside the official survey boundaries are removed in this view. 
During 
preprocessing we then had to remove additional 21 objects where pixels 
contained either infinite numbers or NANs (not a number) in their spectrum. 
Our final sample includes 608\,793 spectra; these are 90\% of the DR4 
spectroscopy main survey.

\subsection{Preprocessing of spectral data}\label{sec:preprocessing}

The preprocessing was performed in 3 steps: 

(1) We reduced the overall size of the data pool to a necessary minimum by 
writing only required data (spectrum, redshift, spectra classification, MJD,
plate id, fiber id) into a single binary file. Other data items from 
the \verb|FITS| file, for instance emission lines, 
continuum-subtracted spectrum, noise in spectrum, mask array, and header 
information were omitted. 

(2) 
The spectra were rebinned to reduce the number of pixels by a factor of 8 and 
the overall file size from 182 KB to 2 KB per spectrum (117 GB to 1.1 GB total). 
This reduction was done by taking the average of two pixels $S_j=(Y_{2j}+Y_{\min(2j+1,n)})/2$ 
for $j=1$ to $n/2$, where $S_j$ is the $j$-th pixel in the smoothed 
spectrum, $Y_j$ the $j$-th pixel in the original spectrum, and $n=3900$ the number of pixels. 
The smoothing was applied iteratively three times over each spectrum.
For the applications discussed in this paper (search of unusual quasars and carbon stars),
the full spectral resolution is not necessary because we are looking for 
unusual continua or broad absorption or emission features which are usually 
at least one order of magnitude broader than the spectral resolution element of the original SDSS spectra. 
Since the SOM algorithm has to
project every single spectrum into a two-dimensional plane only the continuum and the most prominent 
features are considered and several trade-offs have to be made. Indeed the algorithm is very efficient
at this task but it cannot consider every small spectral feature of every input spectrum.  
Therefore, the reduction of the spectral resolution caused by the rebinning
does not significantly reduce the quality of the clustering results as initial tests have shown.  
On the other hand, some applications may require the full spectral resolution. 
One solution would be trading spectral coverage against spectral resolution. 
For instance \citet{Scaringi2009} use a small spectral window from 1401\AA\ to 1700\AA\ for the classification of BALQSOs.

(3) We normalised each spectrum by the total flux density, i.e. the
flux density integrated over the whole spectrum.
To remove gaps of bad pixels that are not marked as OK or emission line 
in the mask array, we used a similar technique as proposed 
by \citet{Jian-qiao01}. 
These gaps were linearly interpolated before the 
reduction process was done.

To mention in passing, we do not transform the spectra into their 
restframes. The main reason is that stars and high-redshift extragalactic objects 
usually share only a narrow restframe wavelength interval;
there is no wavelength overlap at all for quasars with redshift $z \ga 1.5$ and
sources at $z \sim 0$.  Further, the observed spectra are independent of  
wrong redshift determinations from the spectroscopic pipeline. 

\section{Computation of the SOM}

In this section, we describe the generation of the SOM
for about $6\, 10^5$ spectra from the SDSS DR4,
which is a big challenge due to its sheer size. 
The SOM is a very effective algorithm 
that transforms non-linear statistical relationships of the 
original high-dimensional input data (here: spectra) into 
simple geometric relationships in the resulting 
two-dimensional map, which consists of all input 
spectra ordered by their appearance. 

As it is a basic property of SOMs that objects of the same
``spectral type'' tend to form conglomerates and clusters, 
we denote the whole process as ``clustering''.
First, we will briefly describe the basic algorithm and its 
mathematical model; for a full discussion we refer 
to \citet{kohonen1982, SOM} 
from where the mathematical notation was adopted.
Then, in the next section, we discuss in-depth all necessary 
implementation details and considerations taken into account. 

\subsection{The SOM model for spectral clustering}
The set of input variables is defined as vectors 
$\vec{x}(j)=\left[\xi_1(j),..,\xi_n(j)\right]^{T}\in\Re^{n}$ 
where $n=488$ is the number of pixels in each reduced spectrum 
and $j$ denotes the index in the sequence of source spectra 
running from $0$ to $k=608\,792$. 
The neural network then consists of $i\in\left\{1..N\right\}$ neurons, 
represented by weight vectors 
$\vec{m}_i(T)=\left[\mu_{i1}(T),..,\mu_{in}(T)\right]^{T}\in\Re^{n}$, 
that are organised on a two-dimensional grid and $T=0,1,2,..$ 
is the discrete time coordinate. 

\begin{figure}[h]
		\includegraphics[width=0.48\textwidth]{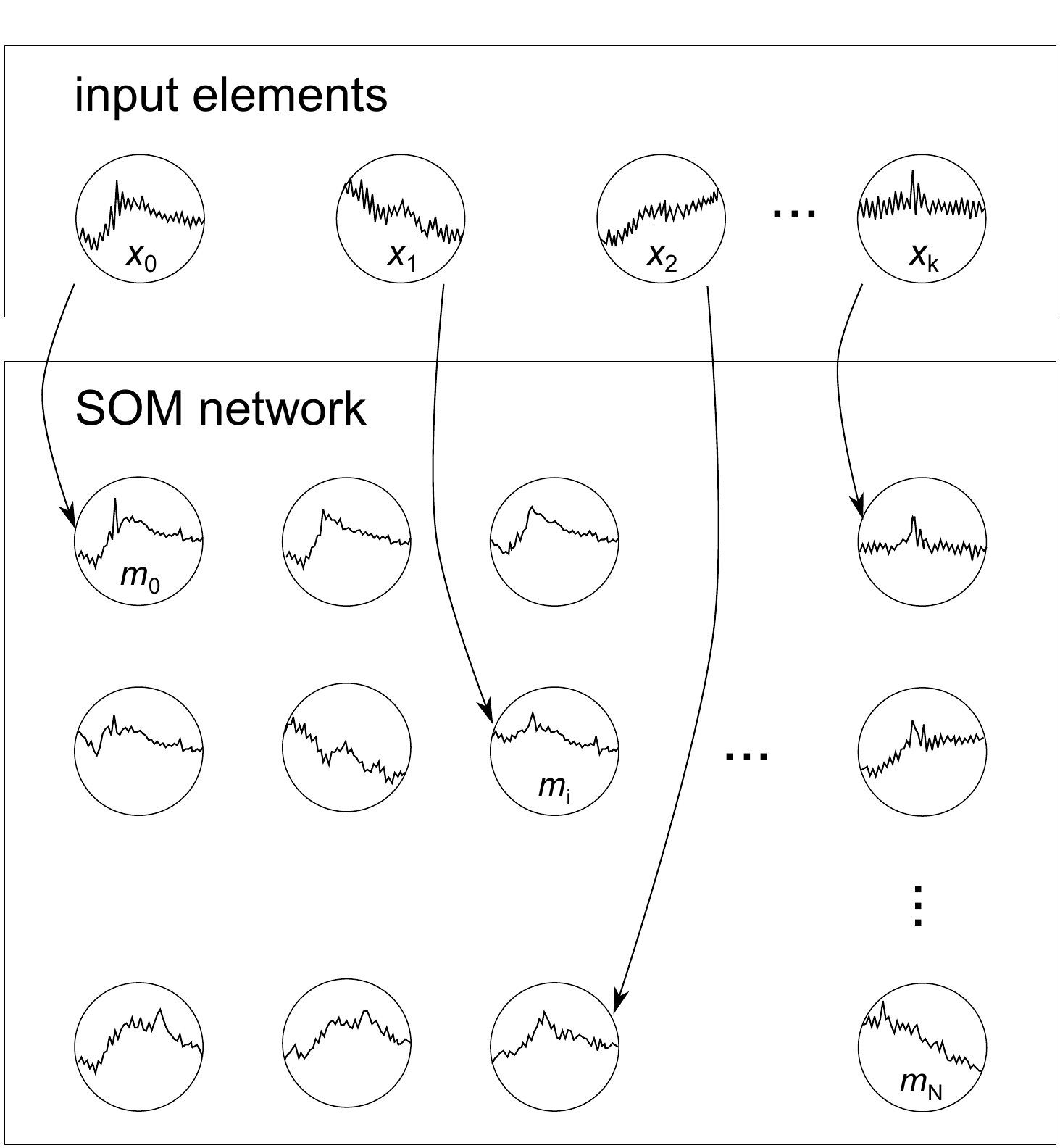}
\caption{SOM network layout: 
The two-dimensional array of neurons $\vec{m}_{\rm i}$. }
\label{fig:nwlayout}
\end{figure}

Typically, neurons are organised on a 
hexagonal lattice. However, 
we have chosen a rectangular lattice, since it allows easier 
and more compact visualisation of our resulting maps as simple 
rectangular images. Regarding boundary conditions a flat grid 
performs best, experiments with cylindrical and toroidal 
topologies reduced the quality of the clustering. 
Figure~\ref{fig:nwlayout} shows the basic network layout with the 
two-dimensional array of neurons $\vec{m}_{\rm i}$. Each 
input element $\vec{x}(j)$ is associated with its best matching 
neuron at every discrete time step $T$. A fraction of neurons 
is empty (has no association with input elements) because $N>k$. 
A detailed discussion about the reasons is postponed to 
Sect.~\ref{sec:nwsize}.  

The process can be initialised by pure randomly chosen weight vectors 
but such an initialisation policy is not the fastest as stated by \citet{SOM}. 
We found that the number of necessary training 
steps is substantially reduced by initialising each weight vector 
$\vec{m}_i(0)$ with a random input spectrum $\vec{x}(j)$. 

The basic SOM algorithm is then based on two important processes 
that are responsible for the self-organising properties of the 
neural network: First choosing a winner neuron $\vec{m}_c$ among 
all $\vec{m}_i$ that has the best match to a given spectrum $\vec{x}$. 
Second, adaption of all neurons in the neighbourhood of $\vec{m}_c$ 
towards $\vec{x}$. 
For each learning step we present each $\vec{x}(j)$ in a random 
order to the network and compute the Euclidean distances 
$\left\|\vec{x}-\vec{m}_i\right\|$ to each neuron $\vec{m}_i$ as a 
measure of dissimilarity. Then, the best matching unit 
(BMU) is defined by the shortest Euclidean distance
\begin{equation}
c=\argmin\left\{\left\|\vec{x}-\vec{m}_i\right\|\right\}.
\label{eq:euclid}
\end{equation}
To prevent collisions in the search for BMUs, where two or more 
different input spectra would share the same neuron, only 
such neurons $\vec{m}_i$ are considered that do not match 
with any of the previously presented input vectors.
The iterated presentation of input vectors in random order over 
many learning steps ensures fairness among all inputs. 
In contrast with a constant sequence, some input vectors would 
receive higher priorities because they appear at the beginning 
of the sequence.

\begin{figure}[h]
 \centering
\includegraphics[width=0.50\textwidth]{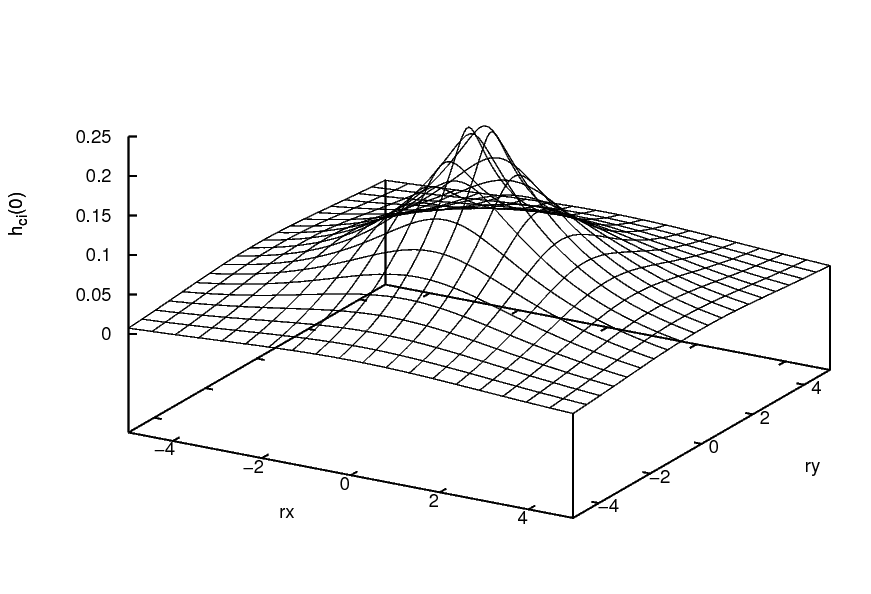}
\caption{
The neighbourhood function $h_{ci}$ at time $t=0$ as a function of the normalised 
radial distances to the BMU, $r_x$ and $r_y$ where the value 1.0 corresponds to 
the map size.
}
\label{fig:hci}
\end{figure}

Then the BMU and all neurons in the neighbourhood are updated 
according to
\begin{equation}
\label{eq:adaption}
\vec{m}_i(T+1)=\vec{m}_i(T)+h_{ci}(t)\big(\vec{x}-\vec{m}_i(T)\big),
\end{equation}
with $t=T/T_{max}$ and where the neighbourhood function 
\begin{equation}
\label{eq:hci}
h_{ci}(t)=\alpha(t)\cdot \exp{\left( -\frac{\left\|\vec{r}_c
-\vec{r}_i\right\|}{2\sigma^2(t)} \right) }
\end{equation}
acts as a smoothing kernel over the network.
With increasing number of learning steps,
$h_{ci}(t)$ approaches zero for convergence. 
Figure~\ref{fig:hci} shows the neighbourhood function 
for the first learning step. $\vec{r}_c \in \Re^2$ is the location 
vector of the BMU and $\vec{r}_i \in \Re^2$ the location vector 
of weight vector $\vec{m}_i$.\\ 
Compared to the frequently used Gaussian kernel, 
our kernel has broader wings and a sharper peak 
at its centre. 
We found from various trials that Eq.\,(\ref{eq:hci}) yields 
better clustering results than its Gaussian counterpart. 
For one-dimensional networks, \citet{Erwin92self-organizingmaps:} 
have shown that convergence 
times are minimal for broad Gaussian neighbourhood functions. 
Employing a function that begins with a large width of the 
order of the largest dimension 
of the network allows rapid 
formation of an ordered map. This is a consequence of 
the absence of metastable stationary states\footnote{States where 
the energy function of the weight vectors, i.e. their change rate, 
reaches a local minimum instead of a global one \citep{Erwin92self-organizingmaps:}.}, 
which slow down 
the convergence progress by orders of magnitudes. 
After an ordered map is formed in the 
first learning steps the width of the kernel can be reduced to 
develop small-scale structures within the map. 

\begin{figure}[h]
\centering
\includegraphics[width=0.50\textwidth]{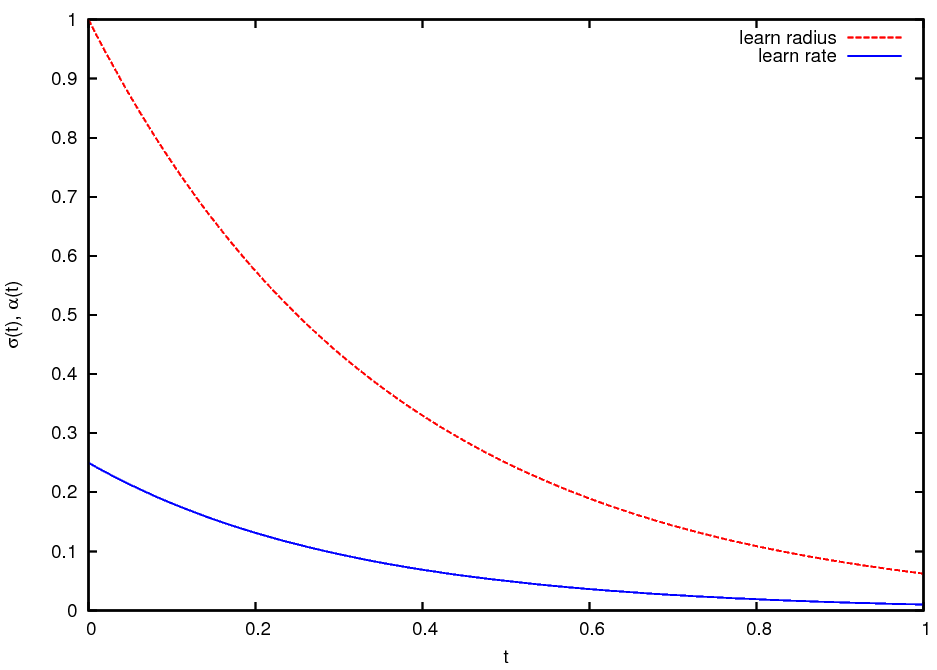}
\caption{Learn radius function $\sigma(t)$ and learn rate function $\alpha(t)$ 
with parameters $\sigma_{\rm begin}=1.0$ $\sigma_{\rm end}=0.0625$, $\alpha_{\rm begin}=0.25$, 
$\alpha_{\rm end}=0.01$.  }
\label{fig:nwparams}
\end{figure}

The neighbourhood function is modified over time by the
learn rate function  
\begin{equation}
\label{eq:alpha}
\alpha(t) = \alpha_{\rm begin}\left( \frac{\alpha_{\rm end}}{\alpha_{\rm begin}} \right)^t 
\end{equation}
and the learn radius function 
\begin{equation}
\label{eq:sigma}
\sigma(t) = \sigma_{\rm begin} \left( \frac{\sigma_{\rm end}}{\sigma_{\rm begin}} \right)^t.
\end{equation}
Both functions are monotonically decreasing over the
time $t = 0 \ldots 1$ altering the neighbourhood function 
in such a way that large-scale structures form in the early training phase 
while small-scale structures and finer details appear at later training steps. 
Figure~\ref{fig:nwparams} shows booth functions for the start and end 
parameters used for the clustering process.

The parameters on the right-hand side of Eqs.\,(\ref{eq:alpha}) and (\ref{eq:sigma})
are the learning parameters of our Kohonen network 
(with $\alpha_{\rm begin} \geq \alpha_{\rm end}$ and 
$\sigma_{\rm begin} \geq \sigma_{\rm end}$). 
In Sect.\,\ref{sec:implementation}, 
we describe a mechanism how those parameters can be chosen properly. 
In order to keep network parameters $\sigma_{\rm begin}$ and $\sigma_{\rm end}$ 
scale-invariant regarding the number of neurons within the network, the 
distance term in $h_{ci}(t)$ should be normalised to the grid size. This 
can be useful when experimenting with different network sizes.

The crucial information of this process is the mapping of input spectra 
to BMUs within the rectangular organised network. After a certain number 
of learning steps, the ordering has taken place and source spectra get 
mapped to the same network location over and over again. Jumps to different 
areas in the map are rare. At this point we obtain the ordered map of input 
spectra as result (see Sect. \ref{sec:number_of_iterations}).

\subsection{Implementation details}\label{sec:implementation}
Before the computation can start, we have to specify all network parameters 
listed in Table~\ref{tab:NWParams}.
Owing to the long computation time of 108 days, it is not possible to 
tweak the network parameters and repeat the entire computation several 
times until a satisfying result in terms of accuracy and convergence 
is reached. Ideally, the clustering of the huge database should be done in 
one shot without successive recomputations. 

\begin{table}[b]
	\caption{Network parameters used for final clustering. }
	\label{tab:NWParams}
	\centering
		\begin{tabular}{l r}                
		\hline                                   
Number of neurons 				$N$ & 859x859 \\
Number of learning steps 		$T_{\rm max}$ & 200 \\
Learn radius 					$\sigma_{\rm begin}$ & 1.0 \\
Learn radius 					$\sigma_{\rm end}$ & 0.0625 \\
Learning rate 					$\alpha_{\rm begin}$ & 0.25 \\
Learning rate 					$\alpha_{\rm end}$ & 0.01 \\
		\hline       
		\end{tabular}
\end{table}

\subsubsection{Deduction of network parameters}
Therefore we deduced all parameters by using a smaller set of artificial 
test ``spectra'' containing sinusoidal signals with increasing 
frequencies $f$ as input data. The limiting frequencies 
$f_{\rm min}$ and $f_{\rm max}$ were chosen arbitrary in a way so that 
oscillation is visible and no aliasing artefacts occur on weight 
vectors $\vec{m}_i$.
This test setting permits to tweak 
all network parameters and shows clearly the goodness of a produced 
clustering. As a success criterion it is required that all test spectra settle finally 
in one coherent structure, sorted by their frequency. 

\begin{figure}[h]
\begin{tabbing}
\includegraphics[width=0.1187\textwidth]{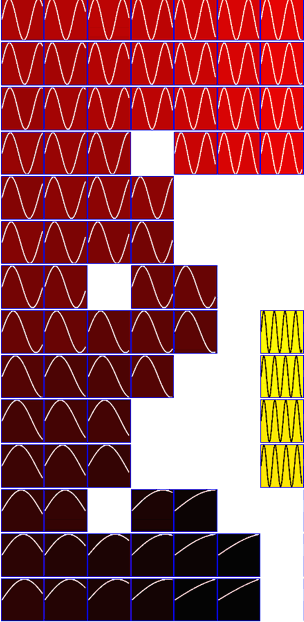}\hspace{-2.2pt}
\includegraphics[width=0.1187\textwidth]{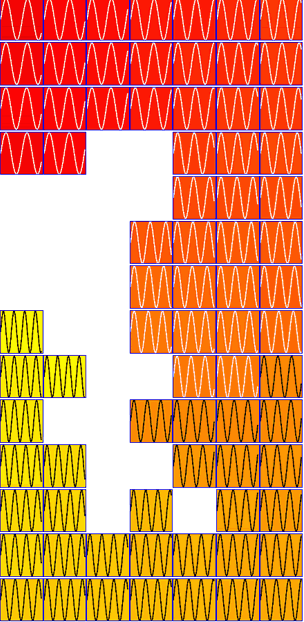}
\includegraphics[width=0.2425\textwidth]{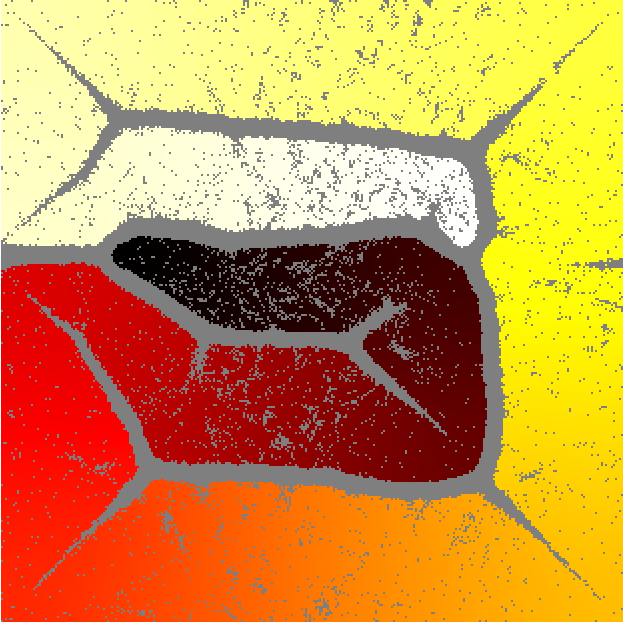}
\end{tabbing}
\caption{
Clustering of sinusoidal test spectra with $N=196$ and $k=150$ (left) 
and $N=96\,721$ (right), respectively.
}
\label{fig:sinetest}
\end{figure}

The best results show a cluster that forms some sort of Hilbert style curve.
The left part of 
Fig.~\ref{fig:sinetest} shows the final clustering result of a 14x14 
map with 140 input elements. 
For validation purposes we repeated this test with the same parameter combination 
for greater sets of test spectra. The right panel of Fig.~\ref{fig:sinetest} shows the clustering 
behaviour of 80\,000 sinusoidal test spectra on a map with 311x311 cells. Empty 
cells are marked grey, frequencies are colour-mapped from black, red, yellow 
to white, where black denotes the lowest frequency.

Experience from many trials with smaller maps and real spectra have 
shown that good clustering results can be achieved with parameter combinations 
that performed well with the ``sinusoidal'' test setting and worse results are 
achieved with parameter combinations that performed poor in the above described test setting. 
However to our knowledge there exists no mathematical proof of the convergence properties 
of the SOM for the general case, i.e. $n$-dimensional input data on a two-dimensional map. 
A proof for the one-dimensional case on an one-dimensional network with a step-neighbourhood function 
was given by \citet{Cottrell1987}, \citet{Cottrell1994} review the theoretical aspects of the SOM. 

\subsubsection{Considerations regarding the size of the network}\label{sec:nwsize}
The number $N$ of neurons in the network must be at least 
equal to the number of 
source spectra in order to guarantee an injective mapping of source spectra. 
However initial tests showed that better results can be achieved if some 
cells are not occupied with source spectra.
For such cells the neurons are not linked to source spectra. 
In the evolution of the neural network, such empty neurons lead to a better 
separation between distinct clusters because they tend to settle at 
the cluster boundaries. The same behaviour is observed for small groups
and even for single outlier spectra.  
Another important factor is the decrease of probability for collisions 
of BMUs when two or more source spectra want to occupy the same neuron. 
Too many empty neurons, on the other hand, (1) scatter similar 
source spectra too much across the map so that no clear cluster boundaries may 
evolve and (2) significantly increase the computing time. A factor of 
$N/k\approx1.2$ produces a good trade-off where similar source spectra 
are not scattered too much
but still have enough room to get into the right clusters.

\subsubsection{Optimisation techniques for faster computations}
We used two optimisation techniques in order to finish the computation 
in a reasonable time frame. The first technique speeds up the search 
phase from $O(N^2)$ up to $O(N)$ for the last learning step. 
For the first learning steps ($T<5$) we conducted a full search 
which requires $\sim k N$ operations per learning step. 
Each operation requires the calculation of the Euclidean distance 
of a source spectrum - weight vector pair.
For all consecutive learning steps, we only searched in the 
neighbourhood of the old winner neuron for each source spectrum $\vec{x}(j)$. 
Since the map is getting more stable with every learning step (due to decreasing $\sigma(t)$) and changes 
are more subtle during the fine-tuning phase, we can lower the search 
radius 
$r_{\rm search}(t)=\left(1-t\right)\sqrt{N}/2+2$ with 
increasing number of learning steps.
The number of operations is then $\sim(1-t)N/4$ per learning step until 
we reach $\sim N$ operations in the last step.

The second technique reduces the number of adaption steps performed by 
Eq.~(\ref{eq:adaption}) by defining a 
threshold. Now the neuron $\vec{m}_i$ is adapted only if the 
neighbourhood function exceeds a predefined value $\tilde{\alpha}$, i.e. 
\begin{equation}
\vec{m}_i(T+1) = \left\{
\begin{array}{ll}
\vec{m}_i(T) & \ {\rm if} \quad h_{ci} \leq \tilde{\alpha} \\
\vec{m}_i(T)+h_{ci}(T)\big[\vec{x}-\vec{m}_i(T)\big] 
             & \ {\rm if} \quad h_{ci} > \tilde{\alpha},
\end{array} \right.
\end{equation}
where we used $\tilde{\alpha} = \alpha_{\rm end}/100$.

\subsubsection{Number of iteration steps and convergence behaviour}\label{sec:number_of_iterations}
\begin{figure}[h]
	\centering
		\includegraphics[width=0.495\textwidth]{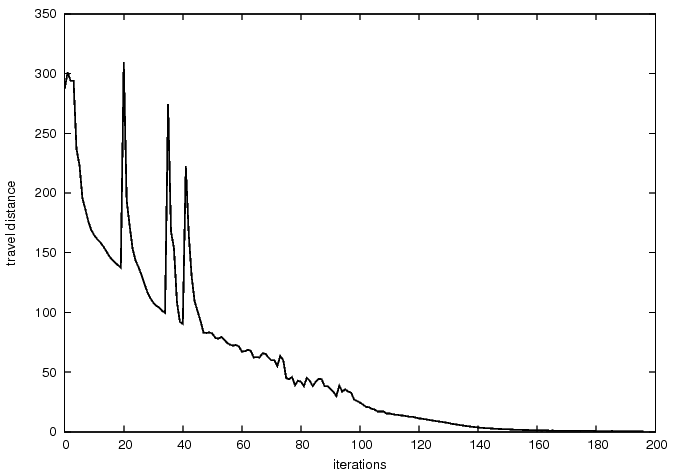}
	\caption{Change in average travel distance (thus the change from one location vector on the map to another) of all source spectra.}
	\label{fig:travel_distance}
\end{figure}

We illustrate the convergence behaviour in two ways. 
First, Fig.~\ref{fig:travel_distance} shows the average travel distance of all 
source spectra. Between each two subsequent learning steps we sum up all location vector changes 
of each source spectrum in the SOM. At certain learning steps, especially in the early 
training phase, major reorganisations within the map occur. Such points can be 
observed in the corresponding visualised maps (presented in the next section) 
at those particular steps.  

Secondly, we calculate 
\begin{equation}
\chi^2(T)=\sum\limits_{j=0}^k\Big(\vec{x}(j)-\vec{m}_{jc}(T)\Big)^2 
\end{equation}
between the source spectra $\vec{x}(j)$ and their corresponding best matching weight vectors $\vec{m}_{jc}$ 
for each learning step $T$. If $\chi^2$ ceases to drop, we can abort the 
learning process at this point. Then the network has reached its optimal point 
between plasticity and stability where the weight vectors still form a smooth landscape. 
We found that the map settles after 200 learning steps. Jumps of source spectra
to different locations are rare in the last learning steps.

\section{Analysis methods}

\subsection{Map visualisation and blending in physical properties}

\subsubsection{Visualisation and presentation of the spectral database} 
After the computation of the SOM finished we built a system that connects all the given information 
and present it in an user-friendly way. 
This system allows the user 
(1) to browse and navigate within the large spectral database, 
(2) to find relations between different objects, 
(3) to search for similar objects from a real or artificial template spectrum.

Each object is represented by an icon that shows its spectrum. 
The background colour encodes the flux density averaged over the spectrum, which can
be used as a proxy for the signal-to-noise ratio in the 
spectrum\footnote{There is a strong correlation between the signal-to-noise ratio and the fiber 
magnitudes. See http://www.sdss.org/dr6/products/spectra/snmagplate.html.
The average flux density in the spectrum, which corresponds to a fiber magnitude 
measured over the whole spectroscopic wavelength window, can thus be used as a 
proxy for the S/N.}. 

\begin{figure*}[tp]
	\centering
\begin{tabbing}
			\includegraphics[width=0.25\textwidth]{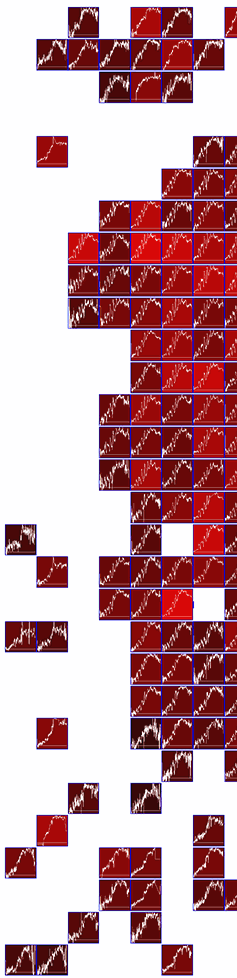}\hspace{-2.2pt}
			\includegraphics[width=0.25\textwidth]{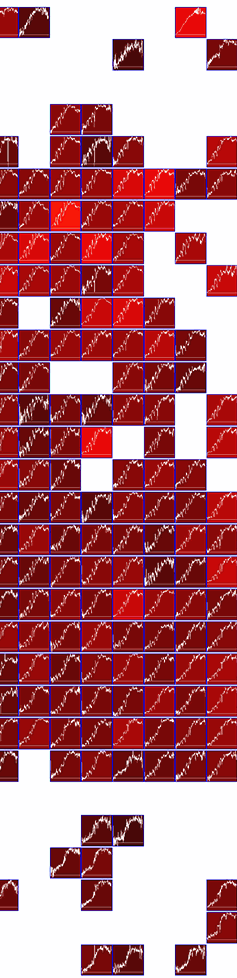}\hspace{-2.2pt}
			\includegraphics[width=0.25\textwidth]{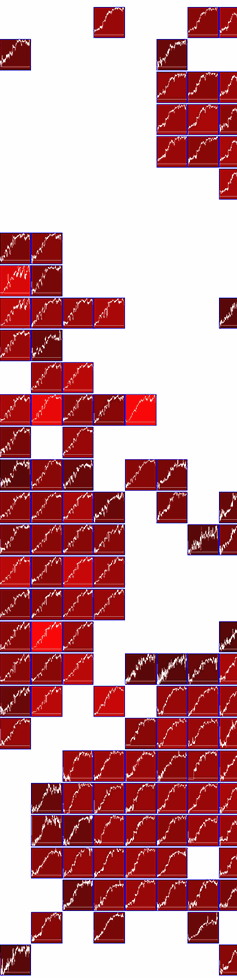}\hspace{-2.2pt}
			\includegraphics[width=0.25\textwidth]{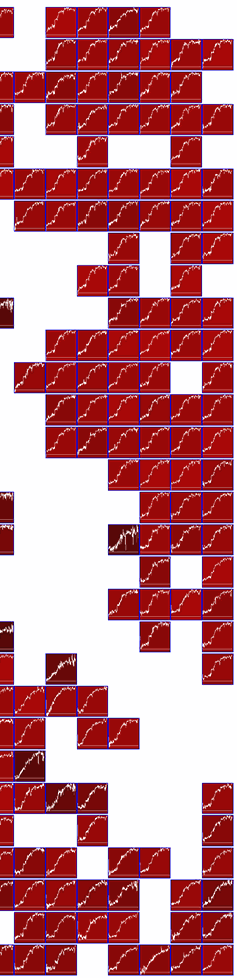}\hspace{-2.2pt}
\end{tabbing}
\caption{Cutout from the icon map including a cluster of carbon stars.}
	\label{fig:sdss_analyze_detailpage}
\end{figure*}

Each object is linked to a summary page that shows the top 20 most similar spectra. 
As similarity measure we use the simple Euclidean distance.
And finally, each object is linked to the 
SDSS SkyServer Object Explorer\footnote{http://skyserver.sdss.org/public/en/tools/explore/obj.asp} 
where additional information can be retrieved. 
Figure~\ref{fig:sdss_analyze_detailpage} displays the blowup of
30x30 spectra from the icon map including a cluster of carbon stars
located in the upper left. White areas show unoccupied cells
without source spectra.

In addition to the icon map, other representations of the SOM are possible:
(1) the difference between the network weights and the corresponding 
input spectra in a logarithmic scale, 
(2) the unified distance matrix (Sect.~\ref{sss:u-matrix}), and 
(3) the $z$ map (Sect.~\ref{sss:phys-prop}) 
using the redshifts from the SDSS spectro pipeline.

We then calculated 
what we call a ``difference map'' for each spectrum. 
The difference map colour codes for each single spectrum in the
SOM its measure of similarity to a given ``template'' spectrum $\vec{y}$ 
which can be either real or artificial as long as it matches the same spectral window and resolution.
Such a map is calculated for every grid cell within the network with

\begin{equation}
d(i) 
= \log\left(\left\|\vec{x}(i)-\vec{y}\right\|+1\right)/
\log\left(\max_j\left\{\left\|\vec{x}(j)-\vec{y}\right\|\right\}+1\right),
\end{equation}
where $\vec{x}(i)$ denotes the spectrum attached to position $i$
in the SOM and $d(i)$ is the difference value in the range $[0,1]$ 
that can be mapped to any colour gradient.
\begin{figure*}[ht]
	\begin{tabbing}

		\includegraphics[width=0.04250\textwidth]{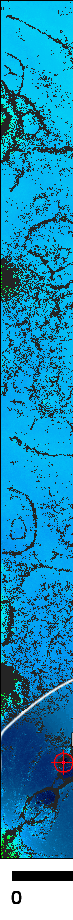}\hspace{-2.2pt}
		\includegraphics[width=0.04133\textwidth]{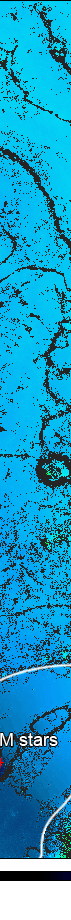}\hspace{-2.2pt}
		\includegraphics[width=0.059416\textwidth]{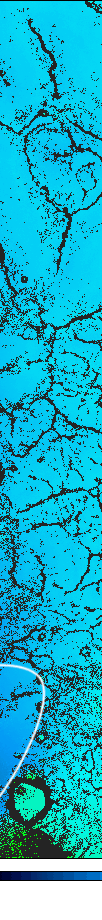}\hspace{-2.2pt}
		\includegraphics[width=0.023883\textwidth]{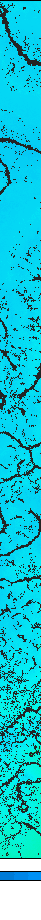}\hspace{-2.2pt}
		\includegraphics[width=0.058251\textwidth]{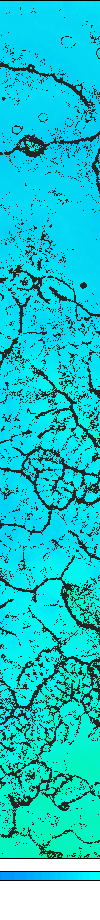}\hspace{-2.2pt}
		\includegraphics[width=0.025048\textwidth]{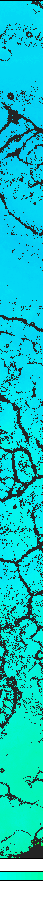}\hspace{-2.2pt}
		\includegraphics[width=0.0833\textwidth]{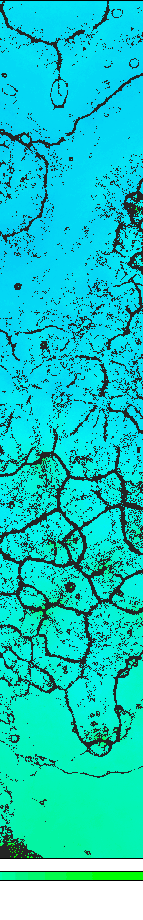}\hspace{-2.2pt}
		\includegraphics[width=0.0833\textwidth]{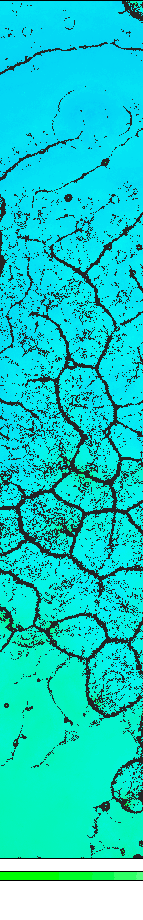}\hspace{-2.2pt}
		\includegraphics[width=0.0833\textwidth]{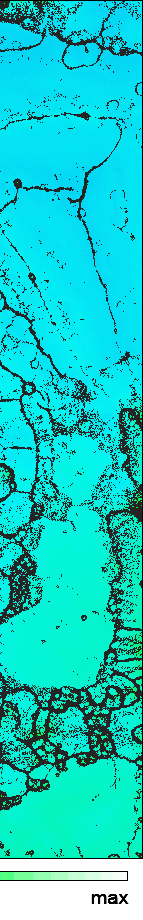}
	
\includegraphics[width=0.05\textwidth]{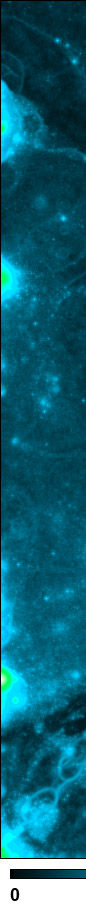}\hspace{-2.2pt}
\includegraphics[width=0.05\textwidth]{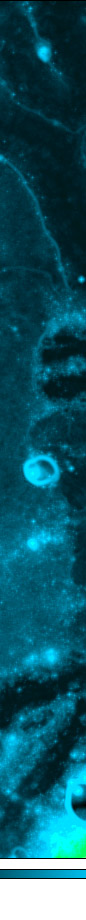}\hspace{-2.2pt}
\includegraphics[width=0.05\textwidth]{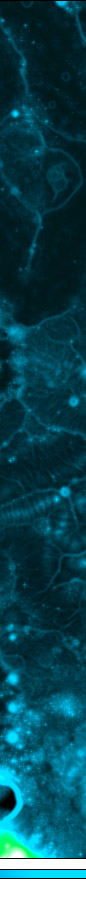}\hspace{-2.2pt}
\includegraphics[width=0.05\textwidth]{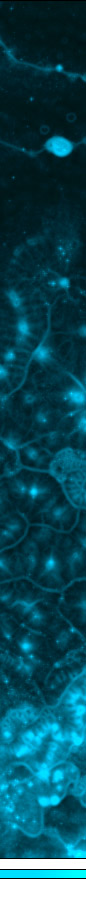}\hspace{-2.2pt}
\includegraphics[width=0.05\textwidth]{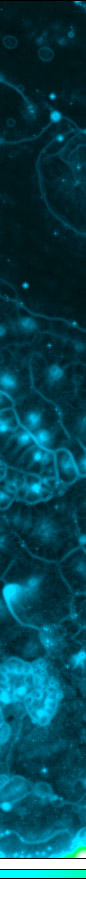}\hspace{-2.2pt}
\includegraphics[width=0.05\textwidth]{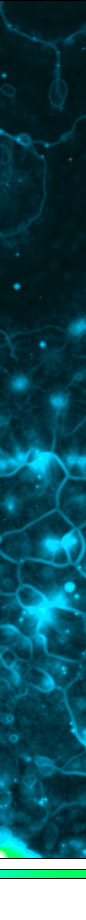}\hspace{-2.2pt}
\includegraphics[width=0.05\textwidth]{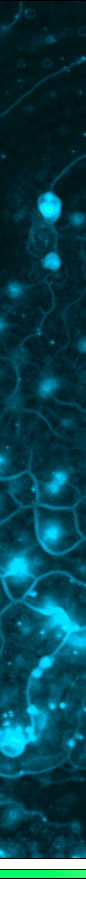}\hspace{-2.2pt}
\includegraphics[width=0.05\textwidth]{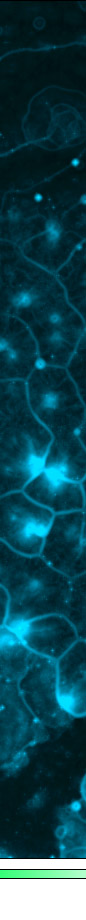}\hspace{-2.2pt}
\includegraphics[width=0.05\textwidth]{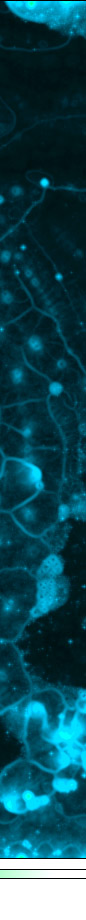}\hspace{-2.2pt}
\includegraphics[width=0.05\textwidth]{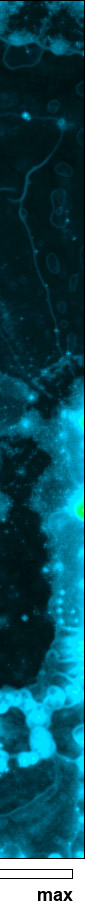}\hspace{-2.2pt}

	\end{tabbing}
	\caption{SOM for $\sim 6\,10^5$ spectra from the SDSS DR4. {\em Left:} Difference map for the M6 star SDSS J092644.26+592553.5. {\em Right:} U matrix of the SOM on logarithmic scale.
}
\label{fig:differencemap_umatrix}
\end{figure*}
For example, Fig.~\ref{fig:differencemap_umatrix} shows the difference map for the
M6 star SDSS J092644.26+592553.5, which is located in the lower left
corner. Such difference maps provide a useful tool to identify objects that
are located in different parts of the SOM, even though their spectral types are similar.
Lighter regions in Fig.~\ref{fig:differencemap_umatrix} show a high degree
of dissimilarity, darker regions show a high degree of similarity.
Grey areas mark free space in the map that is not occupied with spectra.
The dark blue area in the lower left shows an identified cluster of late-type stars.

\begin{figure*}
\centering
\begin{tabbing}
	\includegraphics[width=0.0597767\textwidth]{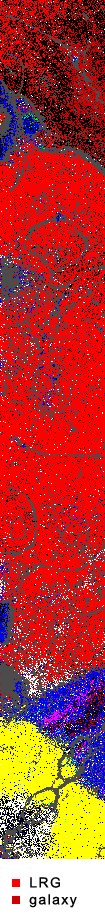}\hspace{-2.2pt}
	\includegraphics[width=0.0626232\textwidth]{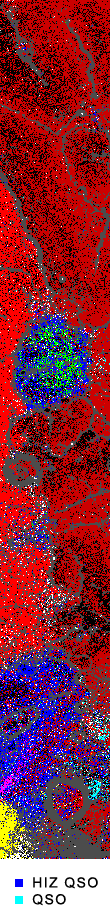}\hspace{-2.2pt}
	\includegraphics[width=0.0609\textwidth]{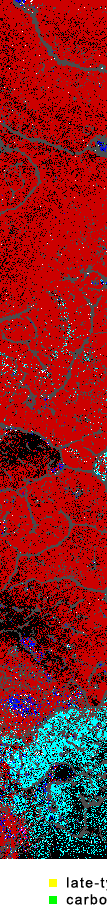}\hspace{-2.2pt}
	\includegraphics[width=0.0626232\textwidth]{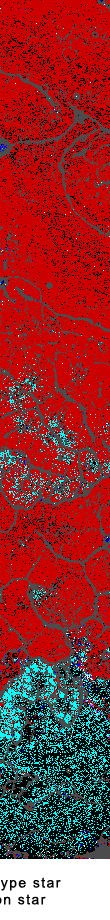}\hspace{-2.2pt}
	\includegraphics[width=0.1224\textwidth]{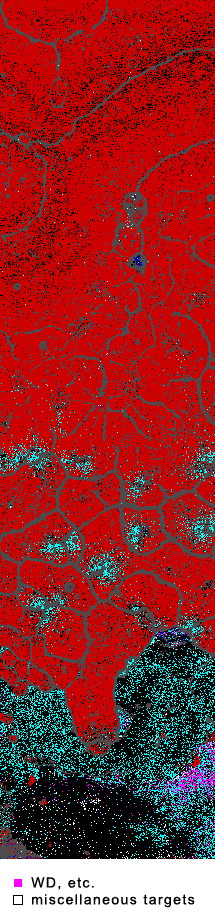}\hspace{-2.2pt}
	\includegraphics[width=0.0597767\textwidth]{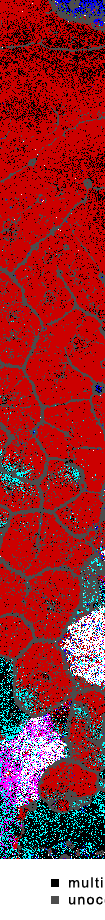}\hspace{-2.2pt}
	\includegraphics[width=0.0632\textwidth]{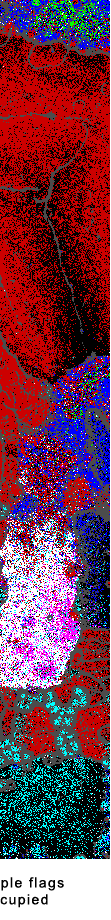}

	\includegraphics[width=0.1224\textwidth]{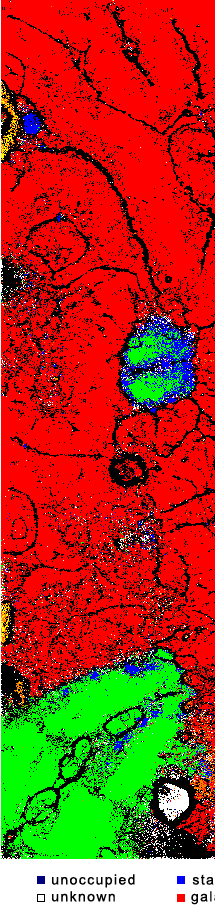}\hspace{-2.2pt}
	\includegraphics[width=0.1224\textwidth]{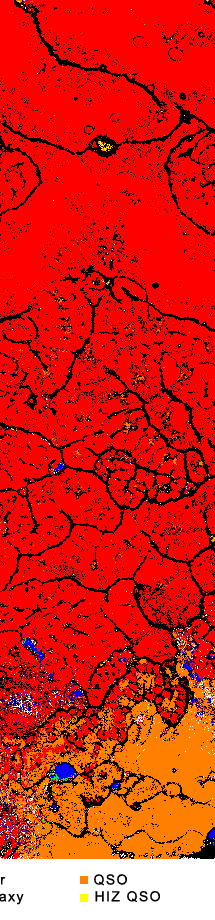}\hspace{-2.2pt}
	\includegraphics[width=0.1224\textwidth]{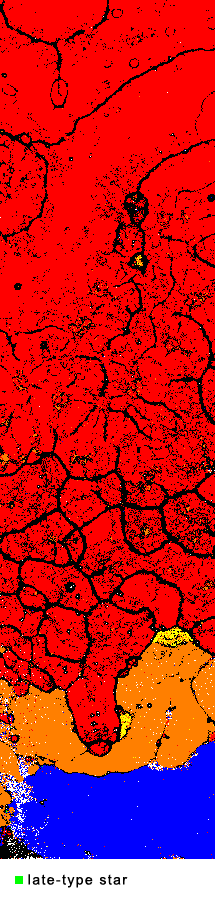}\hspace{-2.2pt}
	\includegraphics[width=0.1224\textwidth]{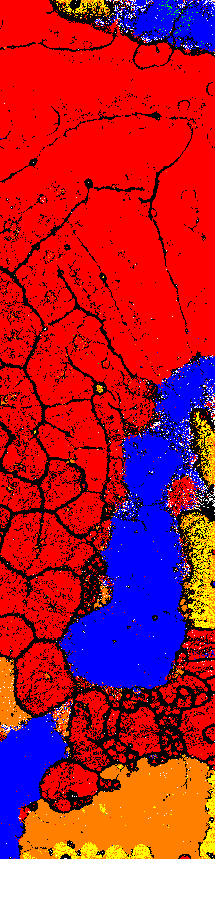}
\end{tabbing}
\caption{The same SOM as in Fig.~\ref{fig:differencemap_umatrix}, but with colour coding representing
the SDSS primary target selection flag ({\em left}) and the classification parameter specClass
resulting from the spectroscopic pipeline ({\em right}).
}
\label{fig:objtype_primTargets}
\end{figure*}

\subsubsection{Unified distance matrix}\label{sss:u-matrix}
The most common visualisation of this particular network is the unified
distance matrix (U matrix) showing the distance between neighbouring neurons 
within the map \citep{ultsch90a}. 
The U matrix is calculated for each weight 
vector $\vec{m}_i$ as the sum of distances of all four immediate neighbours
normalised by the maximum occurring sum of these distances. 
The right panel of Fig.~\ref{fig:differencemap_umatrix} shows the U matrix of the network on a
logarithmic scale at the final learning step.
Lighter colours in the map indicate a high degree of variation,
in contrast darker areas indicate similar weight vectors and clusters
of similar objects. Bigger ``mountains'' (light colours), i.e.
larger distances between neurons, indicate a large
dissimilarity between clusters, smaller mountains indicate similar clusters.

When searching for unusual objects, very small clusters and areas of high 
variation can be of particular interest. The variation is highest at the
cluster boundaries. Boundary regions are usually not occupied with source
spectra because the neuronal landscape changes there from one type to another (see also Fig.~\ref{fig:sinetest}). 
This map is only calculated from the artificial spectra but gives a good 
indication where a lot of change happens, a good indicator to find unusual 
objects.

\subsubsection{Mapping of physical properties}\label{sss:phys-prop}
 
\begin{figure*}[htbp]
\centering
\begin{tabbing}
\includegraphics[width=0.166\textwidth, clip=true]{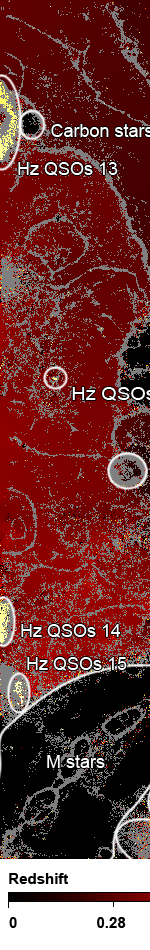}\hspace{-2.2pt}
\includegraphics[width=0.166\textwidth, clip=true]{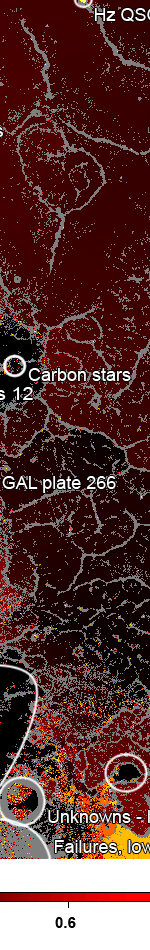}\hspace{-2.2pt}
\includegraphics[width=0.166\textwidth, clip=true]{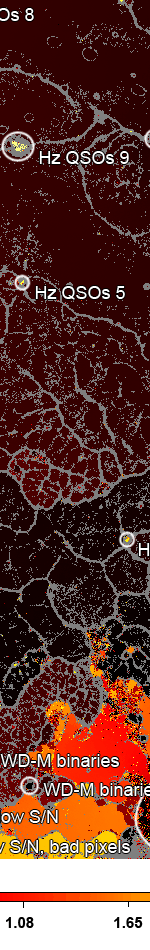}\hspace{-2.2pt}
\includegraphics[width=0.166\textwidth, clip=true]{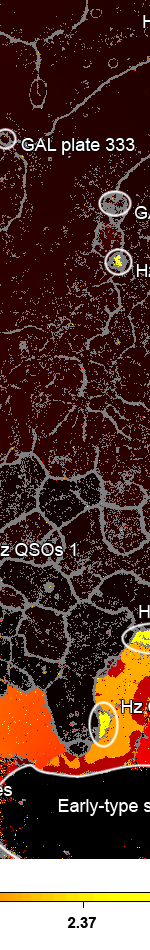}\hspace{-2.2pt}
\includegraphics[width=0.166\textwidth, clip=true]{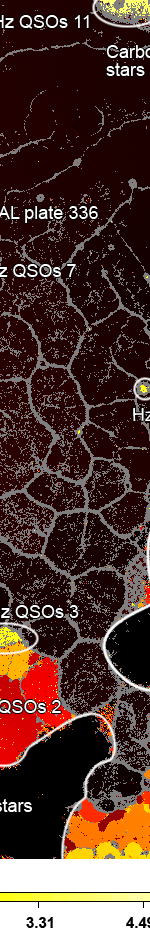}\hspace{-2.2pt}
\includegraphics[width=0.12062\textwidth, clip=true]{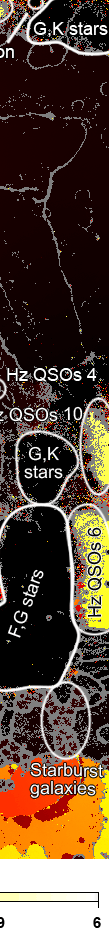}\hspace{-2.2pt}
\end{tabbing}
\caption{The $z$ map with redshifts derived by the SDSS spectroscopic pipeline. Grey areas mark free space
in the map that is not occupied with spectra. We labelled some regions that show
high concentrations of particular objects types.}\label{fig:zmap}
\end{figure*}

In order to gain a deeper understanding of the SOM,
we visualised several physical properties.
In total, we could gather over thirty different maps that describe
various relationships between different
spectral types.
Here we discuss three examples. First, a photometric object classification
parameter is colour-coded. Then, we plot the spectroscopic object classification.
Finally the distribution of the redshift over the SOM is analysed.

The SDSS consists of two surveys, the imaging survey in five specially designed
photometric bands and the spectroscopic survey of objects selected from the
catalogues that were derived from the high-quality five-colour photometry and
the analysis of the image structure. The completely automated algorithm of the
target selection results in a classification of objects as candidates for various
types of galaxies, stars, or quasars. This information is coded in the target selection
flag that is used by the SDSS for the selection of the spectroscopic targets.
In other words, the
target flag stores what that reason was for taking a spectrum. In general,
the ``primary'' selection target bits denote
science targets, and the ``secondary'' target bits denote spectrophotometric standards,
sky targets, and other technical targets. Detailed descriptions of the overall
target selection algorithm are given by
\citet{Stoughton2002SDSSEDR},
\citet{Eisenstein2001SDSSLRG},
\citet{Richards2002SDSSTSQSO}, and
\citet{Strauss2002SDSSTSMGS}.

The left panel of Fig.~\ref{fig:objtype_primTargets} displays the
object classification based on the primary target selection flag.
The colours are attributed to object types as described on the bottom of the panel.
For clarity, several similar object types were combined
(for example, the target flags \verb|QSO_CAP|, \verb|QSO_SKIRT|,
\verb|QSO_FIRST_CAP|, and \verb|QSO_FIRST_SKIRT|
were merged to the type QSO=quasar). HIZ QSO means high-$z$ quasar, LRG means
luminous red galaxy. Objects with multiple target flags are marked black.
The most interesting property of this figure is the clear separation of the
different object types. Within the larger clusters, we observe subtle but continuous
changes in the shape of the continuum and the properties of the emission lines.
Quasar candidates populate a fragmented area at the bottom, but also a number of
isolated clumps scattered across the map. This is to be expected as a consequence of
the wide redshift range covered by the SDSS quasars (see below).

Typically, the parameter \verb|specClass| should be used to characterise the
object type. The class attribute was set by the spectroscopic pipeline
of the SDSS after the spectrum was observed. The following classes are used:
star, late-type star, galaxy, emission line galaxy, quasar (QSO),
high-$z$ quasar (HIZ QSO), and unknown (for unclassifiable spectra). Object
type classification by the SDSS spectroscopic pipeline is discussed
in \citet{Stoughton2002SDSSEDR}.
The visualisation of the class attribute in the right panel of
Fig.~\ref{fig:objtype_primTargets} underlines the separation of object types
in our SOM even stronger than the left panel. An interesting detail
is the strong clustering of the unknown spectral types at the bottom left.
The vast majority of these spectra suffer from a low signal-to-noise ratio.
The lower left corner of the map is populated by late-type stars. The comparison
with the left panel reveals that many of them were targeted as high-$z$
quasars. This is caused by the similarity of the broad-band colours
of these two different object types (see below). 

For an extragalactic survey like SDSS, one of the most interesting visualisations
is the $z$ map that highlights the redshifts $z$ derived by the spectroscopic
pipeline of the SDSS (Fig.~\ref{fig:zmap}). Since the spectra were not
transformed into their  rest-frames, 
a strong ordering and cluster formation towards redshifts can be observed
for galaxies and quasars. We visually inspected a representative number
of spectra from each of the most striking clusters in the SOM to check out
the spectral types. The result is illustrated by the labels in Fig.~\ref{fig:zmap}.

The SDSS quasars cover a redshift interval from $z\sim0$ to $\sim6$ and form several
distinct clusters corresponding to different $z$ intervals. This clustering is a natural
consequence of redshifting the strong emission lines and a demonstration of the colour-$z$
relation of quasars. Quasars with $z \la 2$ populate spatially adjacent areas on the SOM
but also show a clear separation of different $z$ intervals (see the colour bar at the bottom
of Fig.~\ref{fig:zmap}). In addition, we identified 15 separate clusters of high-$z$ quasars
which were labelled in Fig.~\ref{fig:zmap} and listed in Table~\ref{tab:HIZQSO}.
A particularly strong spectral feature is the continuum drop-off shortward of
the Lyman $\alpha$ line at 1216\AA\ (Lyman break) that is caused by the efficient
absorption of UV photons by hydrogen atoms along the line of sight.
The Lyman break enters the SDSS spectral window at $z \ga 2.2$ and moves towards
longer wavelengths with increasing $z$. For redshifts $z\ga4.5$, the continuum is
suppressed by the Lyman $\alpha$ forest shortward of $\lambda \sim 6700$\AA\
and practically completely absorbed by Lyman limit absorption shortward of
$\lambda \sim 5000$\AA. At these redshifts, the optical broad-band colours of the quasars become
similar to those of late-M stars. It is thus not surprising that the highest-$z$ quasars
clump on the SOM in the immediate neighbourhood of the M stars.

\begin{table}[b]
	\caption{High-redshift quasar clusters. }
	\label{tab:HIZQSO}
	\centering
		\begin{tabular}{c c c c c c}                
		\hline\hline                                   
		No. & quantity & $z_{mean}$ & $\sigma$ & $z_{min}$ & $z_{max}$ \\
		\hline  
			1	& 18 & 2.01 & 0.8 & 0.0 & 2.62 \\
			2 & 165 & 2.66 & 0.07 & 1.88 & 2.72 \\ 
			3 & 343 & 2.8 & 0.05 & 2.75 & 2.88 \\ 
			4 & 34 & 2.9 & 0.56 & 0.07 & 3.16 \\ 
			5 & 9 & 3.05 & 0.77 & 0.86 & 3.38 \\
			6 & 2117 & 3.13 & 0.14 & 2.81 & 3.45 \\ 
			7	& 51 & 3.21 & 0.17 & 2.98 & 4.24 \\
			8	& 13 & 3.51 & 0.03 & 2.93 & 3.62 \\
			9 & 65 & 3.82 & 0.86 & 0.0 & 4.32 \\
			10 & 634 & 3.61 & 0.18 & 0.16 & 3.93 \\ 
			11 & 385 & 3.81 & 0.26 & 0.52 & 4.06 \\
			12 & 8 & 3.94 & 0.04 & 3.88 & 4.0 \\
			13 & 344 & 4.06 & 0.19 & 3.53 & 4.42 \\
			14 & 226 & 4.46 & 0.07 & 2.33 & 4.75 \\
			15 & 84 & 4.85 & 0.3 & 3.7 & 5.41 \\			
		\hline       
		\end{tabular}
\end{table}

However the SOM cannot preserve all possible topologies in its two dimensions because 
of the high dimensions of the input spectra. 
A map in three dimensions would allow better arrangements of clusters and more topology information would be preserved. 
On the other hand it would be more difficult to grasp and visualise and may require specialised visualisation software. 
\citet{HGW1994} investigated the dimensionality of input datasets and its effect on topology preservation of the SOM.

\subsection{Tracking of catalogues}
    
\begin{figure*}[hbtp]
\begin{tabbing}
\includegraphics[width=0.2475\textwidth]{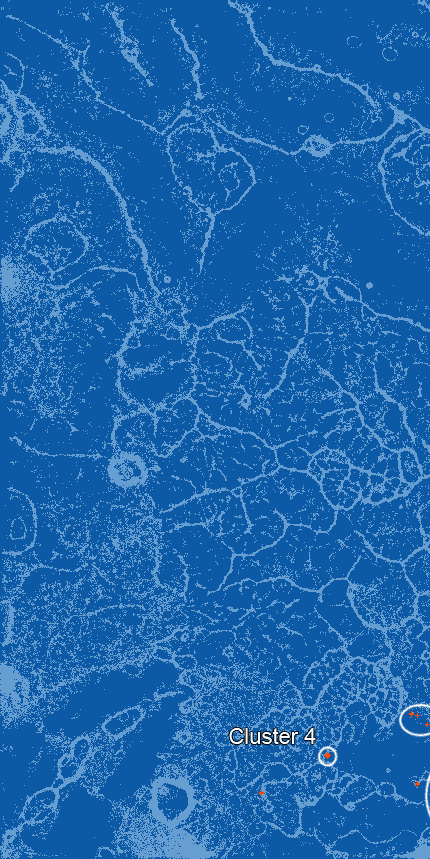}\hspace{-2.2pt}
\includegraphics[width=0.2475\textwidth]{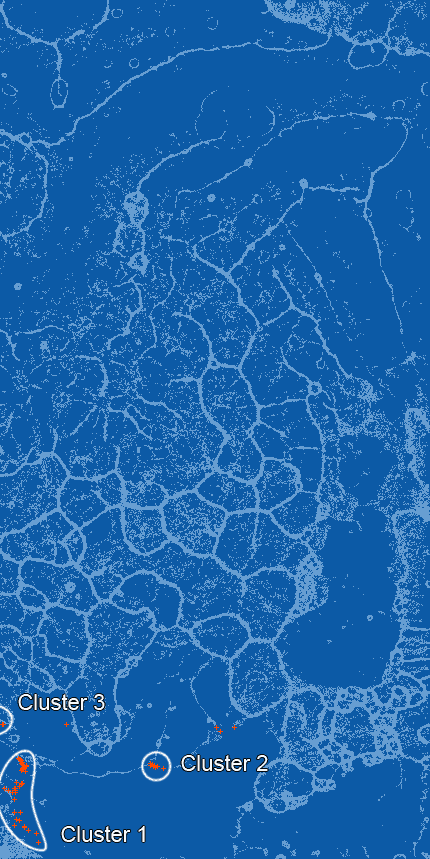}
\includegraphics[width=0.2475\textwidth]{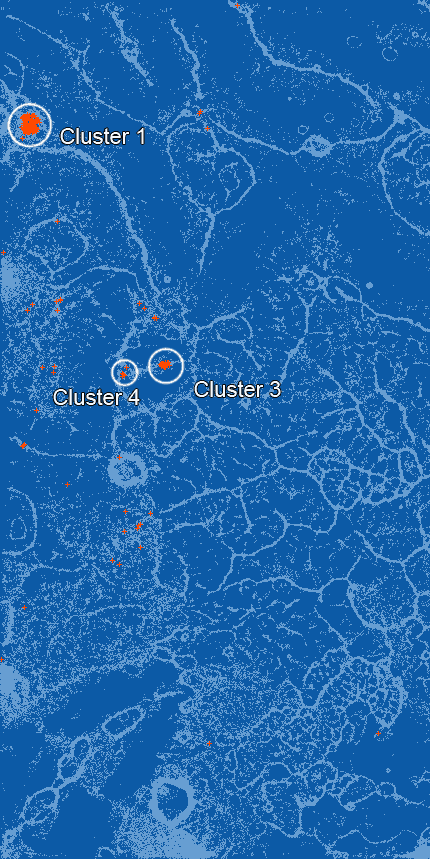}\hspace{-2.2pt}
\includegraphics[width=0.2475\textwidth]{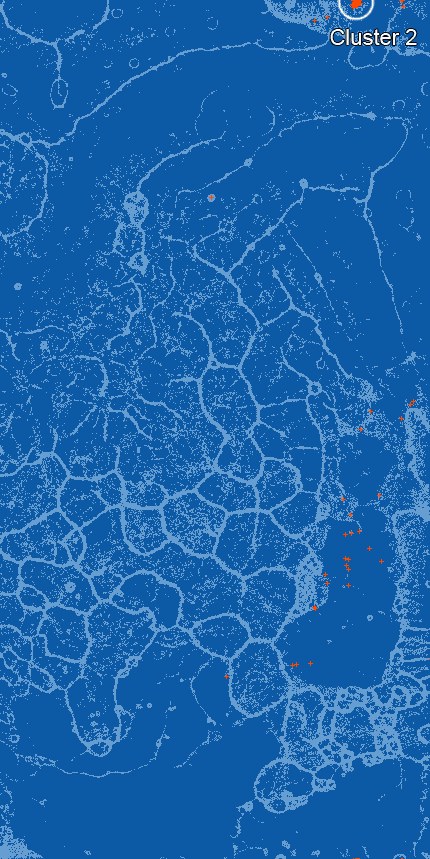}

\end{tabbing}
\caption{SOM object positions and clusters for the white dwarfs of spectral type DQ from \citet{Koester2006}
({\it left}) and the faint high-latitude carbon stars from \citet{Downes2004} ({\it right}).
}
\label{fig:catalogues}
\end{figure*}

For the vast majority of stars, galaxies, and quasars, the spectral properties
vary smoothly over the SOM because stellar spectral types, stellar populations, redshifts,
and dust reddening are continuously distributed in the spectroscopic database of the SDSS.
The bulk of the spectra thus forms large coherent areas interspersed with small
areas of ``no man's land'' occupied either by a mixture of various object types
or by more or less rare types with pronounced spectral peculiarities
(as well as by spectra of low S/N or strongly disturbed spectra). 
If these peculiarities are made of characteristic broad features at fixed
wavelengths in the observer frame, the spectra tend to form small clusters. 
Though it is not easy to specify the relationship between the clustering behaviour 
and the spectral properties, the very fact of such a clustering is useful for 
efficiently searching such rare objects once a cluster has been
identified, e.g. by an input catalogue of known objects of that type.

\subsubsection{Carbon stars}
First, we choose the relatively rare type of carbon stars which display
prominent (Swan) bands of C$_2$ in their spectra.
We use two ``input catalogues'' to trace such objects in the SOM: the catalogue of
65 DQ white dwarfs from Koester \& Knist (2006) and the catalogue
of faint high-latitude carbon (FHLC) stars from \citet{Downes2004}.
The latter catalogue lists 251 C stars of which 231 are
in our database. We are interested how the objects from either catalogue 
are located relative to each other on the SOM. 

A clump of catalogue objects is defined 
to form a cluster if each member is located at a distance $\le 15$ cells from
another cluster member. 
The distribution over the SOM for the objects from the two catalogues is shown in
Fig.~\ref{fig:catalogues} where the four richest clusters are labeled.
The percentage of objects concentrated in the four largest clusters are given
in Table~\ref{tab:catalogues}. Objects that do not fall in one of these clusters 
are listed as \textit{scattered}.

\begin{table}[h]
\caption{Clustering behaviour of catalogued carbon stars. }
\label{tab:catalogues}
\begin{flushleft}
\centering
\begin{tabular}{lrr}
\hline\hline
                                   & DQ    & FHLC \\
                                   & (1)   & (2)  \\
\hline
Total number of objects            & 65    & 231  \\
Percentage of objects in cluster 1 & 58.5  & 45.5 \\
\hspace{2.5cm}... in cluster 2     & 12.3  & 12.5 \\
\hspace{2.5cm}... in cluster 3     &  9.2  &  9.1 \\
\hspace{2.5cm}... in cluster 4     &  6.1  &  2.2 \\
\hspace{2.5cm}... scattered        & 13.9  & 30.7 \\
\hline
\end{tabular}
\end{flushleft}
References. (1)\citet{Downes2004}; (2) \citet{Koester2006}   
\end{table}

\vspace{0.5cm}
\noindent
-- {\it DQ white dwarfs \citep{Koester2006}:}\newline
\indent
White dwarfs of spectral type DQ are defined as showing absorption features 
of carbon atoms or molecules which are believed to be dredged-up from the underlying
carbon/oxygen core to the surface by a deepening helium convection zone. Among others,
DQs are of special interest because they provide information about the deeper layers
of white dwarfs.

The DQ stars are clustered at the borders of the area populated by quasars with 
redshifts around 1. This can be understood primarily as due to their blue continua.
Moreover, the C$_2$ Swan bands resemble broad absorption lines in quasar spectra 
(e.g., SDSS\,J020534.13+215559.7; \citealt{Meusinger2012}), and even broad quasar emission 
lines can be mimicked by the absorption troughs in the case of very strong 
bands.  Though not very compact, the three richest DQ clusters contain 80\% of the 
catalogue objects.

We used the objects from the input catalogue as tracers to search for similar spectra
in their neighbourhood. Since the SOM areas populated by the input catalogue objects
do not show well-defined boundaries, we defined a local neighbourhood around each single
catalogue object by the 8 next neighbours. This yields a list of 365 objects. From the
quick evaluation of the individual spectra we found the following composition of this quite
inhomogeneous mixture of object types:
(1) 153 mostly (93\%) catalogued white dwarfs and 14 catalogued subdwarfs,
(2) 105 extragalactic objects (95 quasars, 4 BL Lac objects, 6 galaxies), and
(3) 93 unclassified, not catalogued objects, mostly (84\%) with featureless blue spectra
(probably DC white dwarfs).
The first group includes 22 DQs from the input catalogue, 19 objects were
found to be classified as DQ by \citet{Eisenstein2006}, another 3 objects are
probably new DQs, yet with only weak und thus uncertain carbon features. 116 objects 
from group 1 are catalogued white dwarfs of other types, mostly DC or DA.

\begin{figure}[htbp]
\centering
\includegraphics[width=0.49\textwidth, trim=40mm 15mm 0mm 10mm, clip=true]{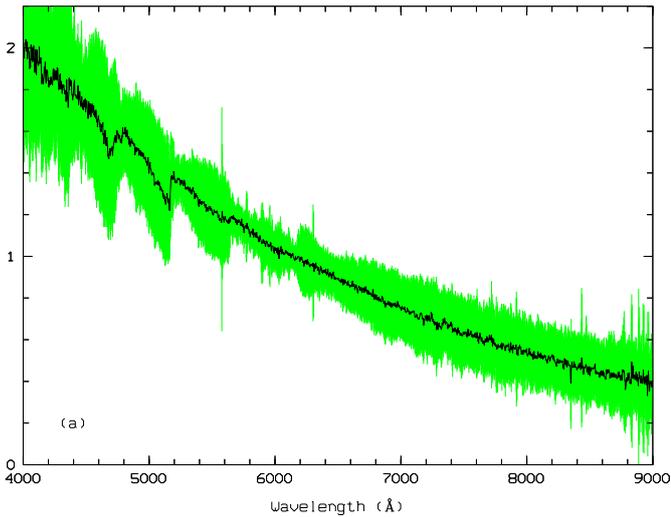}
\includegraphics[width=0.49\textwidth, trim=40mm 15mm 0mm 10mm, clip=true]{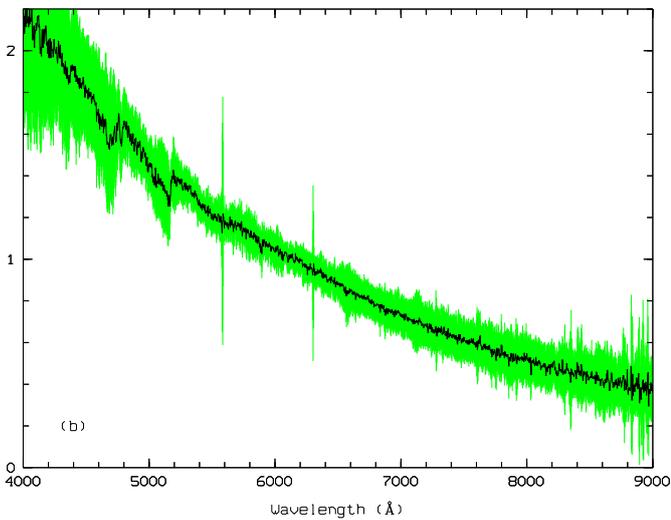}
\caption{
Median and standard deviation for the spectra of DQ white dwarfs
found in the 8-cell neighbourhood of objects from the 
catalogue of \citet{Koester2006}: {\it (a)} for 22 objects which are
in the input catalogue and {\it (b)} for another 22 similar
spectra which are not.
}
\label{fig:DQWDs}
\end{figure}

In Fig.\,\ref{fig:DQWDs}, we compare the median input spectrum with the median spectrum
of the DQ white dwarfs which were ``discovered'' by this method.
This exercise shows that, even for weakly clustering objects of a rare type, new members
can be discovered efficiently by checking the local SOM neighbourhood of known objects.

\vspace{0.3cm}
\noindent
-- {\it Faint high-latitude carbon stars \citet{Downes2004}:}\newline
\indent
FHLCs were considered interesting, among others, as they are believed to be
tracers of the Galactic halo, though recent studies have shown that only a fraction
of them are distant halo giants whereas another significant fraction, maybe the majority,
are nearby dwarfs. The empirical database of the FHLCs has grown substantially
with the SDSS. 

Compared to the DQs, the FHLC stars from \citet{Downes2004} populate
completely different areas of the SOM in the neighbourhood
of intermediate and late-type stars or high-$z$ quasars, respectively.
66\% of the catalogue objects are found to be concentrated in three distinct clusters
with well-defined boundaries. 
There are subtle differences between the mean spectra of the three clumps.
C2 - C3 - C1 form a kind of a spectral sequence where C3 is of later type than
C2 and C1 is of later type than C3. The clusters do not include the stars with the weakest
absorption bands, but most of the stars with very pronounced C$_2$ bands are included,
though some of them are scattered across the map.

\subsubsection{High-Redshift Quasars}
\begin{figure}[htbp]
\centering
\includegraphics[width=0.49\textwidth, trim=40mm 15mm 0mm 10mm, clip=true]{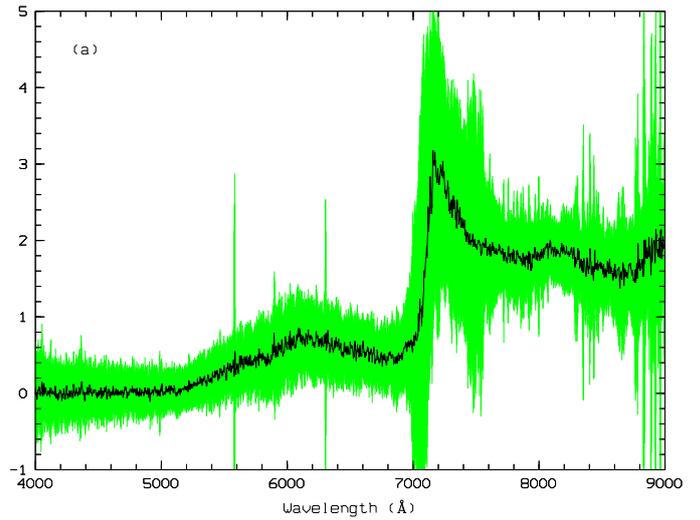}
\includegraphics[width=0.49\textwidth, trim=40mm 15mm 0mm 10mm, clip=true]{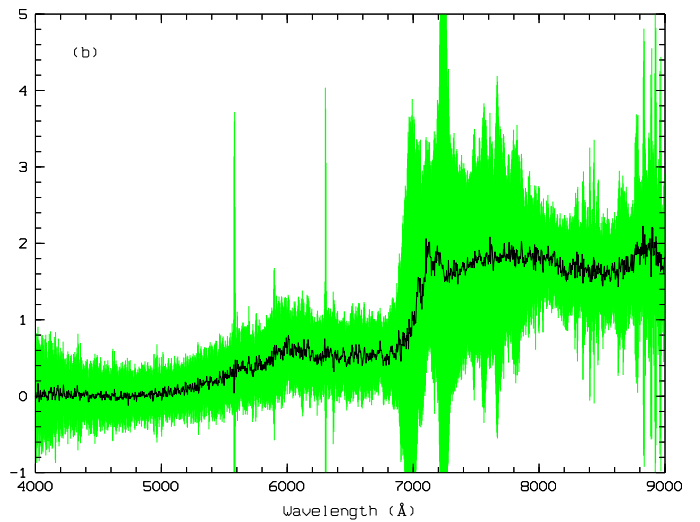}
\caption{
Median and standard deviation for the spectra of the high redshift quasars with $z>4.7$ in cluster 15 {\it (a)}
and outside of cluster 15 {\it (b)}, respectively.
}
\label{fig:hizq15}
\end{figure}

As discussed already in Sect.~\ref{sss:phys-prop}, high-$z$ quasars strongly tend to clump on the SOM.
Here we consider the highest-$z$ quasar cluster 15 (Fig.~\ref{fig:zmap}, Table~\ref{tab:HIZQSO}) for illustration.
This well-defined cluster consists of 84 objects, among them are 78 quasars with $z>4.7$. For
the same redshift range, the SDSS DR7 quasar catalogue \citep{Schneider2010} contains
125 quasars with plate numbers $ \le 1822$, which is the highest plate number in the DR4
database used for our SOM. The completeness of the cluster is thus 62\%, which is somewhat
better than for the biggest clusters of DQWDs and FHLCs, respectively (Table\,\ref{tab:catalogues}).
The fact that more than one third of the highest-$z$ quasars are scattered across the
SOM is not surprising since their spectra can be quite different (Fig.\,\ref{fig:hizq15}).

From the individual inspection of the spectra of all 84 objects we found that
82 spectra are in fact quasars with $z\sim 4$ to 5. Another object, SDSS J153708.14+315854.0,
is likely a galaxy at $z\sim 0.612$, but the S/N in the spectrum is low
and so is the redshift confidence (zConf=0.69).\footnote{The contamination of the high-$z$
quasar cluster with a galaxy of such a low redshift is not unexpected because
the 4000\,\AA\ break of the galaxy spectrum can be easily confused with the
Lyman break when the spectrum is noisy.} For another object, SDSS J084348.13+341255.4,
the red part of the spectrum is so much disturbed that a classification is impossible.
Hence, the search for highest-$z$ quasars in cluster 15 yields a success rate as high as 98.8\%.

\section{Other applications}
\subsection{Quasars}
The advent of large spectroscopic surveys has resulted in
an increase of the number of catalogued quasars by more than one order of magnitude.
The Fifth Edition of the SDSS Quasar Catalogue \citep{Schneider2010} contains 105\,783 entries.
For the vast majority, the individual spectra largely agree with the quasar composite
spectrum produced by averaging over large quasar samples,
i.e. a blue UV/optical continuum and strong broad emission lines. However, these 
surveys revealed also examples of quasars showing dramatically different spectral properties
never seen before, such as very complex systems of absorption features in FeLoBAL quasars 
\citep{Hall2002}, very weak or undetectable UV emission lines \citep{Shemmer2009}, 
extremely red continua \citep{Glikman2007, Urrutia2009a}, or ``mysterious''
objects with spectra that are difficult to explain \citep{Hall2002}. Such rare types 
may be related to special evolutionary stages of the quasar phenomenon and are
expected to shed light on the evolution of active galactic nuclei and their feedback
on the evolution of the host galaxies.     

We started a systematic search for such outliers in the data archive
of about $10^5$ spectra classified as quasars with $z=0.6$ to 4.3
by the spectroscopic pipeline of the SDSS DR7 \citep{Meusinger2012}.
The SOM technique provides us with a unique opportunity for efficiently selecting
rare spectral types from this huge data base. The SOM of the
complete sample is expected to separate the quasars according
to their redshifts (see Fig.\,9). As it was our aim to separate the unusual spectra,
we applied the SOM method to subsamples binned into $z$ intervals.
A bin size of $\Delta z = 0.1$ was chosen to ensure that the
differences between the spectra, as seen by the SOM, caused by their different 
redshifts are smaller than the differences due to the spectral peculiarities. 
The size of the SOMs strongly varies with $z$ between 196 and 8281 neurons.
As outliers tend to settle at the edges and corners of the maps, they were 
easily identified by means of the visual inspection of the icon maps of the 37 SOMs. 
We selected 1530 objects which were individually analysed to reject contaminants
(rare stellar spectral types, spectra with too low S/N, quasars with wrong $z$ from
the SDSS pipeline), to re-estimate the redshift, and to characterise the peculiarities
of the spectra. 

The final catalogue contains 1005 unusual quasars, which could be classified into
6 different types plus a small group of miscellaneous objects.\footnote{The 
spectral atlas for these quasars is available at
http://www.tls-tautenburg.de/research/meus/AGN/Unusual\_quasars.html.
}

Though our catalogue is not complete in a quantifiable sense, it provides the largest
compilation of unusual quasar spectra so far. In particular, the results 
support the idea that these peculiar quasar spectra are not just ``oddballs'', but
represent quasar populations which are probably under\-represented
in the presently available quasar samples.

\subsection{Galaxy Zoo: visualisation of external catalogues}
\begin{figure*}[htbp]
\begin{tabbing}
\includegraphics[width=0.1245\textwidth]{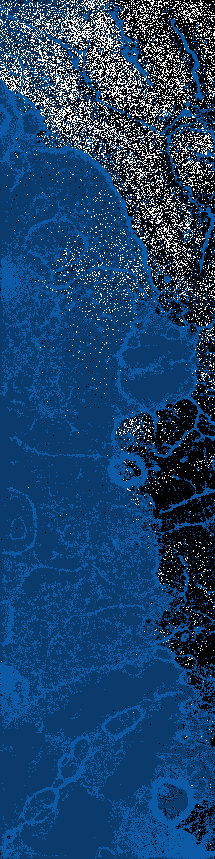}\hspace{-2.2pt}
\includegraphics[width=0.1245\textwidth]{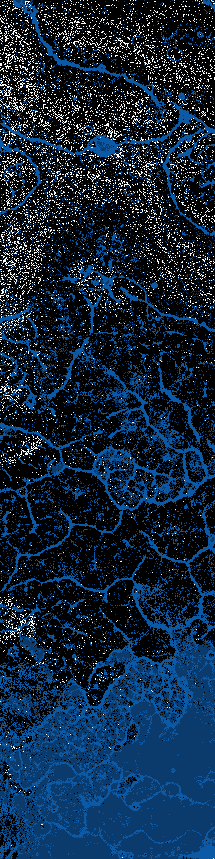}\hspace{-2.2pt}
\includegraphics[width=0.1245\textwidth]{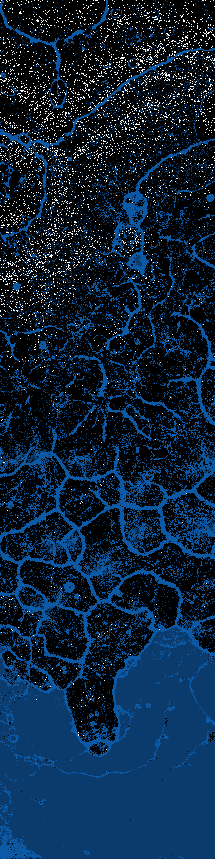}\hspace{-2.2pt}
\includegraphics[width=0.1245\textwidth]{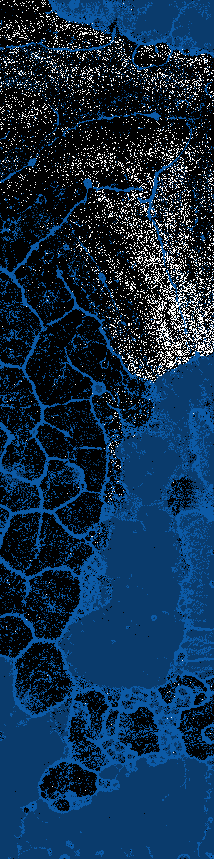}
\includegraphics[width=0.1245\textwidth]{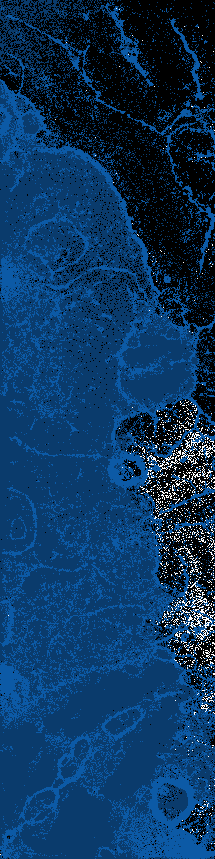}\hspace{-2.2pt} 
\includegraphics[width=0.1245\textwidth]{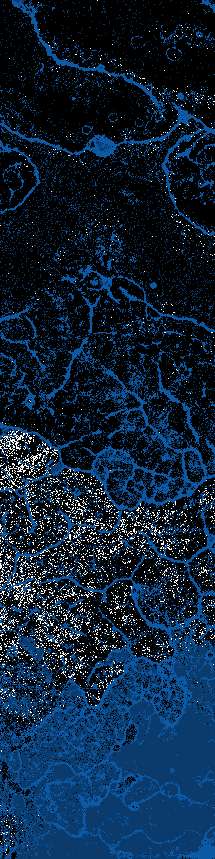}\hspace{-2.2pt} 
\includegraphics[width=0.1245\textwidth]{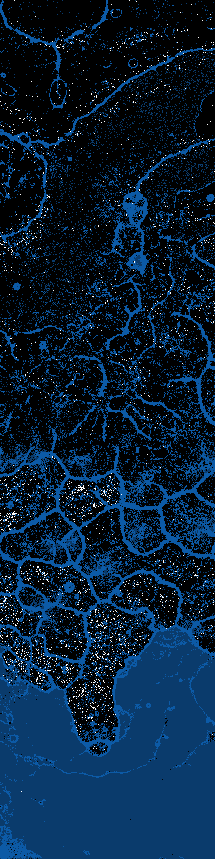}\hspace{-2.2pt} 
\includegraphics[width=0.1245\textwidth]{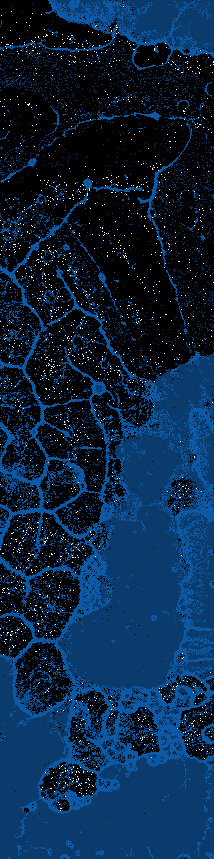}\\
\includegraphics[width=0.1245\textwidth]{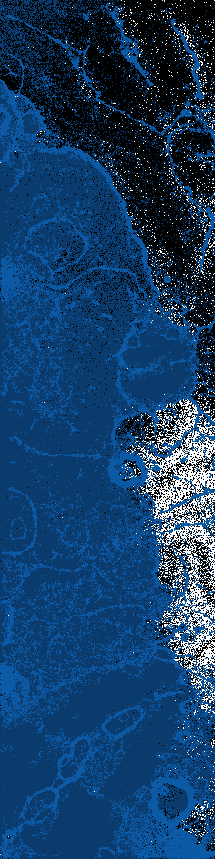}\hspace{-2.2pt}
\includegraphics[width=0.1245\textwidth]{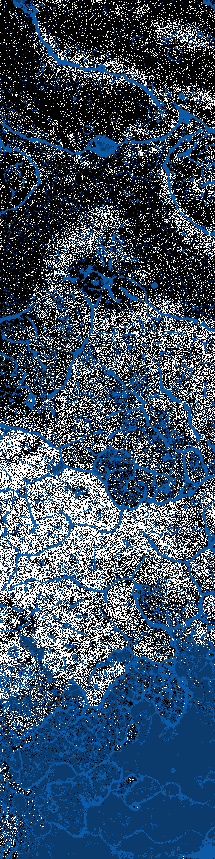}\hspace{-2.2pt}
\includegraphics[width=0.1245\textwidth]{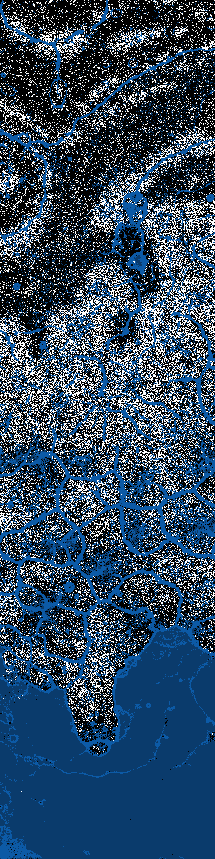}\hspace{-2.2pt}
\includegraphics[width=0.1245\textwidth]{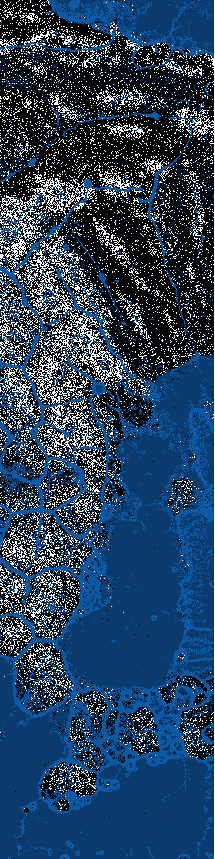}
\includegraphics[width=0.1245\textwidth]{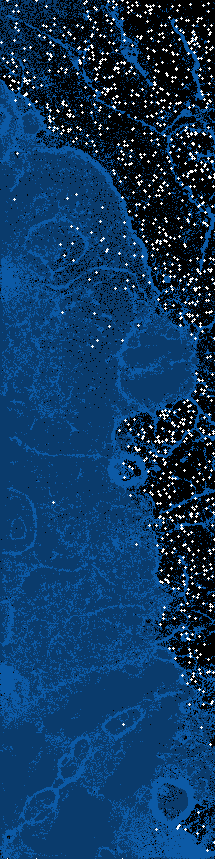}\hspace{-2.2pt}
\includegraphics[width=0.1245\textwidth]{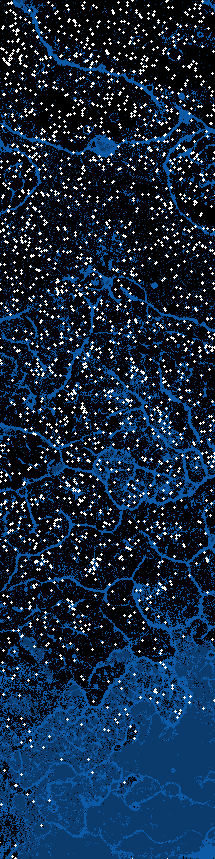}\hspace{-2.2pt}
\includegraphics[width=0.1245\textwidth]{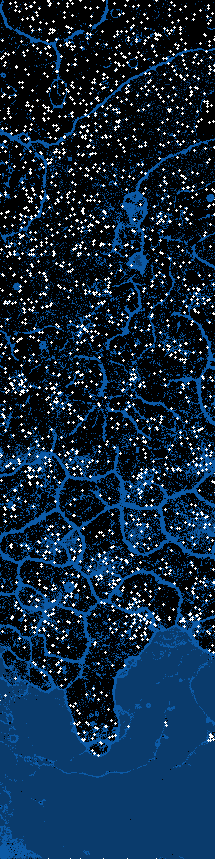}\hspace{-2.2pt}
\includegraphics[width=0.1245\textwidth]{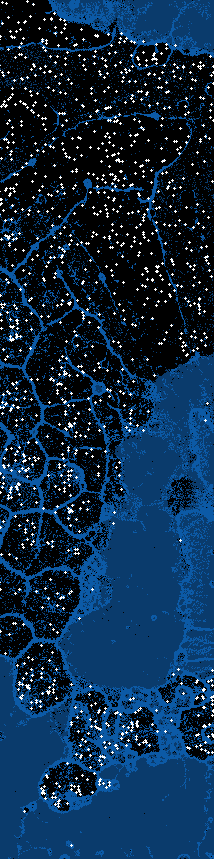}
\end{tabbing}
\caption{The same SOM as in the previous figures where the morphological type flags from the Galaxy Zoo
project \citep{Lintott2011GalaxyZoo} are highlighted as white dots. {\it Left to right then top to bottom:} Elliptical
galaxies, edge-on spirals, spirals (clockwise, anti-clockwise, edge-on), merger.
}
\label{fig:GalaxyZoo}
\end{figure*}

Galaxies account for about three quarters of the SDSS spectra.
For the understanding of galaxies, structure information is crucial and is,
in contrast to quasars and stars, in principle available from the SDSS images.
Galaxy morphology is usually encoded by the morphological type which is a
powerful indicator for the spatial distribution of stars and therewith for
the dynamical evolution of the system, including its merger history.
To gain further insight into the distribution of the galaxies in the SOM,
it may thus be useful to overplot the morphological type information.

Simple morphological classifications were collected by the Galaxy Zoo project
\citep{Lintott2011GalaxyZoo} for 893\,212 objects of SDSS Data Release 6. This
huge project was possible thanks to the involvement of hundreds of thousands volunteer
``citizen scientists''. The galaxies were inspected on composite $gri$ band
SDSS images to derive one of the six classification categories:
(1) elliptical galaxy, (2) clockwise spiral galaxy,
(3) anti-clockwise spiral galaxy, (4) other spiral galaxies (e.g. edge on),
(5) star or Don't know (e.g. artefact), (6) merger.
The results were bias-corrected since faint or/and distant spiral galaxies are
likely misclassified as ellipticals when the spiral arms are not or barely visible
\citep{Bamford2009GalaxyZoo}.

We use the spectroscopically observed subsample of the Galaxy Zoo data. This results
in 667\,945 objects in total and a subsample of 367\,306 objects that overlaps
with the DR4 sample used for our SOM. The catalogue ``Morphological types from Galaxy Zoo 1''
\citep{Lintott2011_2}
lists the fraction of votes for the six classification categories. To turn those
vote fractions into corresponding flags for elliptical or spiral galaxies
requires 80\% of the votes in that category; all other galaxies are classified as uncertain.
For the classification as a merger, a lower threshold of 0.4 is sufficient
(see \citealt{Lintott2011GalaxyZoo}).

Figure~\ref{fig:GalaxyZoo} shows the distribution of the flags
for (left to right then top to bottom) elliptical galaxies (E), edge-on spirals,
spirals (S), and mergers. The redshift increases from right to left on large scales, but there
are deviations on smaller scales. No flags are available for the objects
at the middle of the left edge of the SOM between the M star region and the high-$z$ quasar clusters
14 and 15 at the bottom and the high-$z$ quasar cluster 13 at the top (see Fig.\,\ref{fig:zmap}).
This region is occupied by the highest-$z$ galaxies where 
morphological information from the SDSS imaging is not reliable for the vast majority
of the galaxies. Nearly all flagged galaxies within that area were assigned to type E.

A few interesting details can be recognised by the simple inspection of Fig.~\ref{fig:GalaxyZoo}.
First,  E galaxies populate mostly the upper part of the SOM, the type S is concentrated
towards the lower half. However, there are no clear boundaries between the areas populated by
E and S galaxies, respectively. In particular, the region in the middle of the upper part (around
12 o'clock) is populated by comparable fractions of E and S galaxies. On smaller scales, however,
the two types are stronger separated in a kind of a meshwork structure. The cracks running through
the high-density S area between about 7 and 9 o'clock are loosely populated by spectra of
E-type galaxies and surrounded by a remarkable concentration of edge-on spirals. Finally,
the spectra of merger galaxies do obviously not show a preference for any morphological type.
Also, there is no enhanced population density of mergers in the clump of starburst galaxies
labelled in Fig.\,\ref{fig:GalaxyZoo}.
A few loose clumps of mergers are found at the boundary between galaxies and
intermediate-redshift quasars, representing wet mergers with elevated
star formation producing blue continua and strong emission lines. The detailed investigation of
these issues is clearly beyond the scope of the present paper.

\subsection{Galaxy maps from SDSS DR7}
\begin{figure}[htbp]
\begin{tabbing}
\includegraphics[width=0.2425\textwidth]{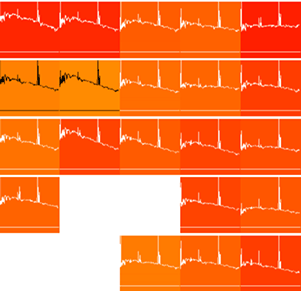}
\includegraphics[width=0.2425\textwidth]{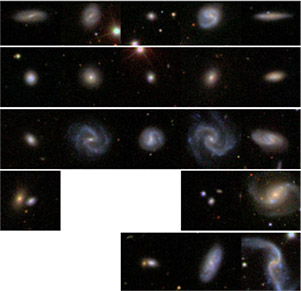}
\end{tabbing}
\caption{Image cutouts from an icon map (left) and the corresponding picture map (right) of low-redshift galaxies.}
\label{fig:PictureMaps}
\end{figure}

We finally note in passing that we computed SOMs for $\sim 8\,10^5$
galaxies from the SDSS DR7. As for the quasars in the previous subsection,
the galaxies were binned into $z$ intervals with about 5000 spectra per bin.
The analysis of the results is still in preparation. An additional powerful tool
for the work with the resulting SOMs are picture maps (Fig.~\ref{fig:PictureMaps}), i.e. representations of the
SOMs where the colour images from the SDSS are displayed at the positions of the
corresponding spectra. The comparison of the icon maps with the picture maps are
expected to be helpful when searching for correlations between spectral properties and
morphology or environment.

\section{Conclusions and future work} 
In this paper we have presented ASPECT, a software tool that is able 
to cluster large quantities of spectra with the help of self-organising maps 
(SOMs; \citealt{kohonen1982, SOM}).
We have built a topological map of 608\,793 spectra from the SDSS DR4 database
to illustrate the capability of that tool.
To explore the resulting topology information we have created a system that links 
each spectrum in the map to the SDSS DR7 explorer. ASPECT allows the user
to browse and navigate through the entire spectral data set.
Similarities within the SOM have been visualised with the help of the unified distance matrix \citep{ultsch90a}.
Further we have introduced difference maps that colour code the similarity of a given template spectrum to all 
other spectra in the SOM. Data from different sources
were mapped onto the resulting SOM.
Especially the mapping of SDSS photometric and spectroscopic object types (Fig.~\ref{fig:objtype_primTargets}) and SDSS 
derived redshifts (Fig.~\ref{fig:zmap}) onto the resulting SOM enable a better navigation within the data set.

Clusters of rare objects within the SOM can be identified either by the visual inspection of selected spectra or with the
help of a given input catalogue of known objects of that type.
The first method has been successfully applied for selecting unusual quasars from $10^5$ SDSS DR7 spectra in our
previous study \citep{Meusinger2012}. Here we demonstrate the second method by means of
65 DQ white dwarfs from \citet{Koester2006} and
231 Faint high-latitude carbon stars from \citet{Downes2004}.
From those catalogue objects 86\% DQ white dwarfs and 69\% FHLCs are concentrated in four major clusters respectively.
By checking the SOM neighbourhood of those clusters similar objects can be discovered efficiently, even for weakly clustering objects.
As another application we have mapped morphological information (i.e. galaxy types, mergers) from the Galaxy
Zoo project onto the spectroscopic galaxy subsample of the SOM. As shown in Fig.~\ref{fig:GalaxyZoo}, elliptical galaxies,
spirals, and edge-on spirals show different distributions across the map. Merger galaxies, on the other hand,
do not show a preference for any morphological type.
More detailed galaxy morphology information, for example from the Galaxy Zoo2 project which data release is currently
prepared but not yet available, is expected to offer interesting results when mapped to the here presented topology.

Data mining of other existing or upcoming massive spectroscopic surveys for instance the Sloan Extension
for Galactic Understanding and Exploration \citep[SEGUE;][]{Yanny2009SEGUEStars} or Apache Point Observatory
Galactic Evolution Experiment (APOGEE) would offer great potential.

Further challenges involve the overcome of algorithmic limitations (runtime and memory bandwidth usage)
of the current algorithm used in ASPECT. A distribution of the workload on modern supercomputers
would enable the processing of even larger data sets. The source code is available on request for the interested reader.

\begin{acknowledgements}  
We thank our anonymous referee and Dr. Polina Kondratieva for their important comments and suggestions.  
  
This research would be impossible without the use of data products from Sloan Digital Sky Survey (SDSS). 
Funding for the Sloan Digital Sky Survey (SDSS) has been provided by the Alfred P. Sloan Foundation, the Participating Institutions, the National Aeronautics and Space Administration, the National Science Foundation, the U.S. Department of Energy, the Japanese Monbukagakusho, and the Max Planck Society. The SDSS Web site is http://www.sdss.org/. 
The SDSS is managed by the Astrophysical Research Consortium (ARC) for the Participating Institutions. The Participating Institutions are The University of Chicago, Fermilab, the Institute for Advanced Study, the Japan Participation Group, The Johns Hopkins University, the Korean Scientist Group, Los Alamos National Laboratory, the Max-Planck-Institute for Astronomy (MPIA), the Max-Planck-Institute for Astrophysics (MPA), New Mexico State University, University of Pittsburgh, University of Portsmouth, Princeton University, the United States Naval Observatory, and the University of Washington.

This research has made use of the SIMBAD database, operated at CDS, Strasbourg, France and of NASA's Astrophysics Data System Bibliographic Services. 
\end{acknowledgements}

\vspace{3cm} 

\bibliographystyle{aa}
\bibliography{publ}

\end{document}